\documentclass[traditabstract]{aa}
\usepackage{txfonts}
\usepackage{natbib}
\usepackage{graphicx}
\usepackage{flushend}
\usepackage{afterpage}
\usepackage{appendix}
\usepackage{lscape}
\usepackage{psfrag}
\usepackage{threeparttable}
\usepackage{microtype}
\usepackage{float}
\usepackage{subfigure}
\usepackage{afterpage}
\usepackage{multirow}
\usepackage{longtable}

\def\apjl{ApJL}

\def\aap{A\&A}
\def\apjs{ApJS}

\begin{document}

   \title{The role of cold gas and environment on the stellar mass - metallicity relation of nearby galaxies}

   \subtitle{ }

   \author{T. M. Hughes\inst{1},
   	  L. Cortese\inst{2},
	  A. Boselli\inst{3},
	  G. Gavazzi\inst{4},
      J. I. Davies\inst{5}}

\institute{Kavli Institute for Astronomy \& Astrophysics, Peking University, Beijing 100871, P.R. China
                \email{tmhughes@pku.edu.cn}
           \and
	      	European Southern Observatory, Karl-Schwarzschild Str. 2, 85748 Garching bei Muenchen, Germany	
                \email{lcortese@eso.org}
                \and
		Laboratoire d'Astrophysique de Marseille, UMR6110 CNRS, 38 rue F. Joliot-Curie, F-13388 Marseille France
              \email{Alessandro.Boselli@oamp.fr}
		\and
		Universita' di Milano-Bicocca, piazza della Scienza 3, 20100, Milano, Italy
             \email{giuseppe.gavazzi@mib.infn.it}
             \and
		School of Physics and Astronomy, Cardiff University, Cardiff, CF24 3AA, UK 			\email{Jonathan.Davies@astro.cf.ac.uk}
}

   \date{Accepted for publication in A\&A}

\newcommand{\hi}{H{\sc i}} 
\newcommand{\hii}{H{\sc ii}\ }
\newcommand{\oi}{O{\sc i}}
\newcommand{\oii}{O{\sc ii}}
\newcommand{\hd}{H{\sc $\delta$}}
\newcommand{\hg}{H{\sc $\gamma$}}
\newcommand{\hb}{H{\sc $\beta$}}
\newcommand{\oiii}{O{\sc iii}}
\newcommand{\oiv}{O{\sc iv}}
\newcommand{\nii}{N{\sc ii}}
\newcommand{\niii}{N{\sc iii}}
\newcommand{\ha}{H{\sc $\alpha$}}
\newcommand{\sii}{S{\sc ii}}
\newcommand{\siii}{S{\sc iii}}
\newcommand{\cone}{C$_{1}$}
\newcommand{\rz}{R$_{23}$}
\newcommand{\oz}{O$_{32}$}
\newcommand{\zzz}{12+log(O/H)}
\newcommand{\neii}{Ne{\sc ii}}
\newcommand{\mgii}{Mg{\sc ii}}
\newcommand{\mercury}{Hg{\sc i}}
\newcommand{\kms}{km~s$^{-1}$\ }

  \abstract{
We investigate the relationship between stellar mass, metallicity and gas content for a magnitude- and volume-limited sample of 260 nearby late-type galaxies in different environments, from isolated galaxies to Virgo cluster members. We derive oxygen abundance estimates using new integrated, drift-scan optical spectroscopy and the base metallicity calibrations of Kewley \& Ellison (2008). Combining these measurements with ultraviolet to near-infrared photometry and \hi \ 21 cm line observations, we examine the relations between stellar mass, metallicity, gas mass fraction and star formation rate. We find that, at fixed stellar mass, galaxies with lower gas fractions typically also possess higher oxygen abundances. We also observe a relationship between gas fraction and metal content,
whereby gas-poor galaxies are typically more metal-rich, and
demonstrate that the removal of gas from the outskirts of spirals
increases the observed average metallicity by $\sim$0.1 dex. Although some cluster galaxies are gas-deficient objects, statistically the stellar-mass metallicity relation is nearly invariant to the environment, in agreement with recent studies. These results indicate that internal evolutionary processes, rather than environmental effects, play a key role in shaping the stellar mass-metallicity relation. In addition, we present metallicity estimates based on observations of 478 nearby galaxies.}


	\authorrunning{T. M. Hughes et al.}
	\titlerunning{The role of cold gas and environment on the stellar mass-metallicity relation}
   \maketitle
 
\section{Introduction}

The chemical composition of galaxies provides a crucial insight into the processes governing galaxy evolution. There are still many open questions about the processes that drive metal enrichment and the importance of the role of the environment in shaping the evolution of galaxies. As galaxies evolve, star formation converts gas into stars, which in turn produce the heavy elements via nucleosynthesis. These metals are then expelled into the surrounding medium during the later stages of stellar evolution, thus enriching gas that may become fuel for future star formation episodes. Therefore the abundance of heavier elements present in a galaxy, the metallicity, provides an important indicator of the evolutionary history of a galaxy.

Since the discovery of a relationship between luminosity and metallicity by \citet{leq1979}, numerous studies have confirmed the existence of a luminosity-metallicity or stellar mass-metallicity (hereafter M-Z) relation (see e.g. \citealp{rubin84}; \citealp{skillman89}; \citealp{vila92}; \citealp{garnett2002}). Stellar mass and metallicity both trace the integrated star formation history of a galaxy. The gas-phase metallicity also probes feedback processes affecting the chemical enrichment of the interstellar medium. Therefore, since the evolutionary stage of a galaxy can be inferred from knowledge of these two quantities, the M-Z relation provides a valuable tool for studying the chemical evolution of galaxies. Yet, despite mounting observational evidence for the existence of the M-Z relation, many questions remain regarding the origin, scatter and the possibility of a dependence on the environment. 

\citet[hereafter T04]{tremonti2004} investigated the local M-Z relation for 53,000 local galaxies, finding that the
M-Z relation is steep for masses $\leq$ 10$^{10.5}$ M$_{\odot}$ and flattens at higher
masses. With the use of chemical evolution models, they interpret the flattening in terms of efficient galactic winds that remove metals
from low-mass galaxies ($\leq$ 10$^{10.5}$ M$_{\odot}$). However, these observations can also be explained if low-mass galaxies possess low star formation efficiencies, caused by supernova feedback \citep{brooks2007}, or via a variable integrated stellar initial mass function (IMF; \citealp{koppen2007}).

In addition to these various interpretations for the origin of the relation, there is also debate about the impact of the environment on the M-Z relation. Many early works based on small samples of nearby galaxies (e.g. \citealp{ssk1991}; \citealp{hplc1992}; \citealp{hpc1994}; \citealp{skill1996}) suggest that galaxies residing in the closest cluster of galaxies, the Virgo cluster, typically display metallicity enhancements up to 0.2 dex greater than similar objects in sparser environments, suggesting that the M-Z relation may vary with environment. However, more recent studies of larger samples of galaxies have found that the effect of the environment is much weaker than previously claimed. \cite{mouhcine2007} selected a sample of over 37000 galaxies from the SDSS and examined the dependence of the oxygen abundance on the stellar mass and local density. Across a range of environments, from isolated systems to the periphery of clusters, they reported only a weak dependence of the M-Z relations on environment, with changes in metallicity between 0.02 and 0.08 dex occurring over a factor 100 change in local density. They also found that for a fixed stellar mass, galaxies in denser environments only display a slight metal enhancement compared to galaxies in sparser environments. Moreover, \cite{ellison2009} found that the metallicity enhancement in the cluster M-Z relation is up to 0.05 dex, warning that environmental differences are subtle and may not be clearly observed in the unbinned data of even large samples ($>$1300) of galaxies. In contrast, \cite{cooper2008} find a stronger relationship between metallicity and environment for galaxies in their SDSS-based sample, linking $\geq$ 15\% of the scatter in the observed M-Z relation to the environment. Thus, there appears to be some disagreement about whether or not the environment plays a significant role in the chemical evolution of galaxies.

Before we can properly tackle these open issues, we must first address a more pressing problem; the various methods used to estimate metallicities often produce highly discrepant results. Gas-phase oxygen abundances, a proxy of metallicity, are determined from observations of nebular emission lines in optical galaxy spectra. The [\oiii] $\mathrm{\lambda}$4363 line may be used to directly determine the oxygen abundance (the ‘direct’ or ‘T$_{e}$’ method), but detection of the line requires long integration times, making large surveys unfeasible. Instead, observations of large samples of giant local extragalactic \hii regions, where the [\oiii] $\mathrm{\lambda}$4363 line is more easily detected, provide empirical relationships between different optical emission line ratios and oxygen abundances determined via the T$_{e}$ method. These calibrations can then be applied to galaxy spectra to estimate the oxygen abundance. In addition to these empirical metallicity calibrations, theoretical calibrations using nebular photoionization models (e.g. CLOUDY, \citealp{ferland98}) combined with stellar population synthesis models (e.g. Starburst99, \citealp{starburst99})  allow the theoretical emission line ratios to be predicted from varying input metallicities, temperatures and densities. Many empirical and theoretical calibrations have been developed that utilise different suites of emission lines, such as [\oiii] $\mathrm{\lambda}$3727, [\oiii] $\mathrm{\lambda}$4959, [\oiii] $\mathrm{\lambda}$5007 and \hb \ in the R$_{23}$ method (e.g. \citealp{p79}; \citealp{m91}; \citealp{kk04}; \citealp{p05}), the [\nii] $\mathrm{\lambda}$6584/\ha \ ratio (e.g. \citealp{d02}; \citealp{pp04}) and the [\nii] $\mathrm{\lambda}$6584 / [\oii] $\mathrm{\lambda}$3727 ratio (\citealp{kd02}). Another approach is to simultaneously fit all the strong emission lines and use theoretical models to generate a probability distribution of metallicities and statistically estimate the abundances (T04). 

Despite the plethora of calibrations available, there is little agreement between the estimated metallicities. \cite{liang06} compared the metallicity calibrations from four different methods applied to $\sim$ 40000 galaxies selected from the SDSS \citep{sdss2000}, and observed discrepancies as large as 0.6 dex between observational estimates and the results from photoionization models. \cite{yin2007} also found a discrepancy of 0.6 dex when comparing the theoretical results of T04 to the empirical calibration of \citet{p01}. These large discrepancies in the metallicities obtained from different calibrations mean that obtaining reliable methods for measuring the metallicity of a galaxy is still a focus of current research efforts. 

This problem is important since hierarchical galaxy formation models within the $\Lambda$CDM framework which incorporate chemical evolution models are now able to predict the evolution of the M-Z relation (see e.g. \citealp{delucia2004}; \citealp{derossi2007}; \citealp{dave2007}; \citealp{bertone2007}). Reliable observations are required in order to accurately determine the shape of the M-Z relation, and thus constrain current theories of both the formation of galaxies and their subsequent chemical evolution. Advancements in the capabilities of telescopes and spectrographs allow for the observation of the M-Z relation out to higher redshifts (e.g. \citealp{kobul1999}; \citealp{erb2006}; \citealp{liang06}; \citealp{mouhcine2006}). However, observations at low and high redshifts cover different restframe wavelengths, meaning that different emission lines are used to determine the metallicity. Since the discrepancies between metallicities from different calibrations may be as large as the expected evolution, these systematic errors in the calibrations need to be first understood, so that the observed M-Z relations at higher redshifts may accurately constrain the models.

Motivated by the need for an in-depth study of these discrepancies, \citet[hereafter KE08]{ke08} compared M-Z relations derived using ten different calibrations from the literature. In addition to finding large systematic discrepancies of up to 0.7 dex in the metallicities from different calibrations, there was an evident disparity in the overall shape and zero point of the M-Z relations (see Fig. 2 of KE08). Such discrepancies mean that comparing results of different studies is not possible and it is crucial to use the same metallicity calibration for any comparisons of the M-Z relation. False trends in the data may also be introduced in studies that try to maximise the number of available measurements by using a combination of different calibrations. In an effort to facilitate comparisons between the results of various samples, KE08 determined conversions allowing metallicities derived with different calibrations to be converted into the same base calibration. The new conversions removed the 0.7 dex systematic discrepancies. Using these new metallicity conversions, it is now possible to compare metallicities derived using different calibrations reliant on different sets of emission lines. A better estimate of the relative oxygen abundance of a galaxy may be obtained by converting each metallicity estimate into the same base metallicity and then using the average metallicity from all the available estimates. This method has the advantage of maximising the amount of spectral information used to estimate the metallicity. With more reliable estimates, it may be possible to accurately determine the properties of the M-Z relation and potentially explore the different theories of chemical evolution which give rise to the relation. 

In this paper, we first test the metallicity calibrations and base conversions presented in KE08 with new integrated, drift-scan optical spectroscopic observations of galaxies in the Herschel Reference Survey \citep{boselli2010}. For the first time, we study the relationships between stellar mass, metallicity and gas content using direct measurements of atomic hydrogen. Throughout this work we adopt the \textit{oxygen abundance} as a tracer of the overall gas phase \textit{metallicity}, using the two terms interchangeably, and a solar oxygen abundance of \zzz = 8.69 (\citealp{asplund2009}).

\section{The samples and data}

\begin{figure}[ht]
\begin{center}
\includegraphics[width=0.45\textwidth]{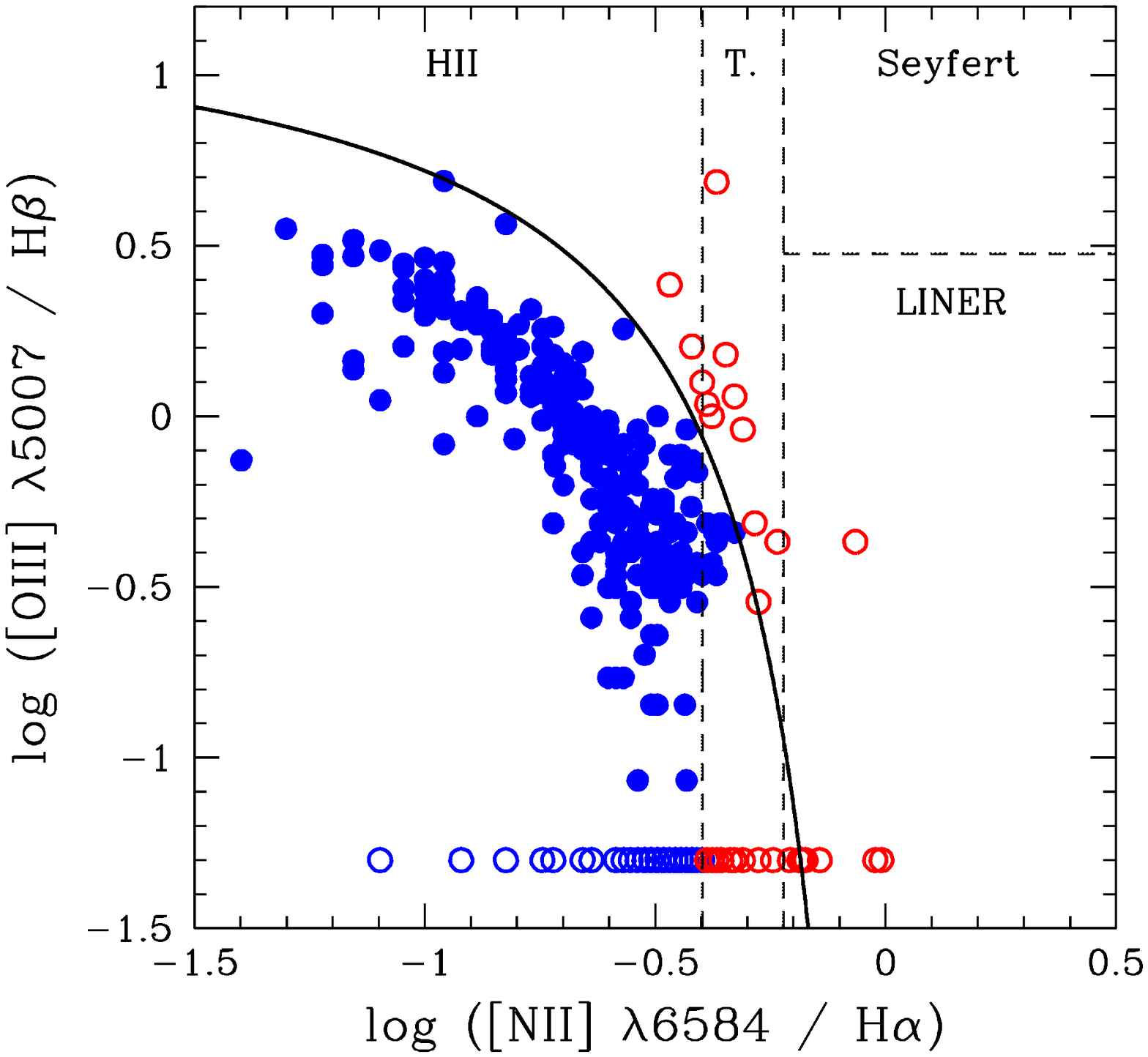}
\end{center}\vspace{-0.3cm}
\caption[The BPT diagnostic diagram used to classify the spectra of galaxies in this work]{The BPT (\citealp{bpt81}) diagnostic diagram used to classify the spectra of galaxies in this work, based on the [\nii]/\ha \ and [\oiii]/\hb \ line ratios. Objects for which a measurement of [\oiii]/\hb \ is not available are plotted at log([\oiii]/\hb) = -1.3.  The \citet{decarli2007} scheme (straight dashed lines) marks the classification boundaries between star-forming, LINER, Seyfert and composite/transition galaxies. In addition, the demarcation line between star-forming and AGN-host galaxies from \citet[solid black line]{kauffmann2003} is shown. Potential AGN-host galaxies (red open circles) are thus removed from the sample of normal star-forming systems (blue closed circles).}\label{fig:bpt}
\end{figure}

In this study, we adopt a galaxy sample drawn from late-type galaxies in the Herschel Reference Survey \citep[hereafter HRS]{boselli2010}. The HRS consists of a volume- and magnitude-limited sample, selecting late-type (Sa and later) galaxies within a distance range of 15 $\le Dist. \le$ 25 Mpc and a 2MASS
\citep{skrutskie2006} K-band magnitude K$_{Stot} \le 12$ mag.
Finally, galaxies residing at high galactic latitude ($b >$ +55$^{\circ}$) and in regions suffering from low galactic extinction ($A_{B}$ $<$ 0.2; \citealp{schlegel1998}) are selected to minimize galactic cirrus contamination. These criteria produce a sample of 260 late-type galaxies. As extensively discussed in \citet{boselli2010}, this sample is
not only representative of the local universe but also spans different density regimes (i.e., from isolated systems to the center
of the Virgo cluster) and so it is ideal for environmental studies (see also \citealp{hughes2009}; \citealp{ch2009}). For the following analysis, two sub-samples are created based on the environment inhabited by each object in the sample. All galaxies with membership of the Virgo cluster, as determined with the criteria defined in \citet{gav1999}, comprise the `Virgo' sub-sample. Galaxies outside Virgo, which range from isolated systems to galaxies in groups, are selected for the `non-Virgo' or `field' sub-sample (and we refer to this sample of galaxies in low density environments using both terms interchangeably).

A large multi-wavelength dataset is available for the HRS sample. For estimating oxygen abundances, we use flux measurements of optical emission lines presented in \citeauthor{boselliprep} (2012), which discusses the data in detail. To briefly summarise, observations were carried out from 2004-2009 using the CARELEC spectrograph on the 1.93 m telescope at Observatoire de Haute Provence. Galaxies were observed using a 5 arcmin slit of width of 2.5 arcsec, adapted for the typical seeing conditions (2-3 arcsec), with a typical wavelength range of 3600-7000 \AA \ incident on the CCD at a resolution of R$\sim$1000. Integrated spectra were obtained for 135 HRS galaxies by observing in the drift-scan mode, whereby the spectrograph slit is allowed to drift across the galaxy major axis, and the data reduced using standard tasks in the IRAF\footnote{IRAF is the Image Analysis and Reduction Facility made available to the astronomical community by the National Optical Astronomy Observatories, which are operated by AURA, Inc., under contract with
the U.S. National Science Foundation. STSDAS is distributed by the Space Telescope Science Institute, which is operated by the Association of Universities for Research in Astronomy (AURA), Inc., under NASA contract NAS 5-26555.} software package. By combining this data with emission line measurements from integrated spectra presented in \citet{gavazzi2004}, \citet{kennicutt1992}, \citet{jansen2000}, \citet{moustakas2006} and \citet{moustakas2010}, a homogeneous dataset of the spectral properties of 236 of the 260 HRS late-types has been assembled. The spectra are corrected for internal and galactic extinction and the \ha \ and \hb \ lines are corrected for underlying stellar absorption. For \hb \ lines with detected absorption, the absorption feature is deblended from the emission line using SPLOT. To be consistent with Gavazzi et al. (2004), a mean additive correction of 1.8 in flux and -1.4 \AA \ in E.W. is applied to those \hb \ lines where underlying absorption is not detected. For \ha \ lines, the spectral resolution and close proximity of the [\nii] lines prevent the measurement of the \ha \ underlying absorption. However, it has been shown that, despite quite different star formation histories, the underlying absorption of the \ha \ line is fairly constant with a mean value of E.W.H$\alpha_{abs}$ = 2.80 $\pm$ 0.38 \AA \ ~(Moustakas \& Kennicutt 2006). We thus apply a standard correction of 2.80 \AA \ in E.W. and

\begin{equation}
{f(H\alpha)_{corr} = f(H\alpha)_{obs} \times (1 + \frac{2.8}{E.W.H\alpha_{obs}})}
\end{equation}

\noindent
for the flux for all galaxies. The emission line intensities are then corrected for internal and galactic extinction using the Balmer decrement $C(H\beta)$ and the reddenning function of \citet{fitz2007}, as described in Boselli et al. (2012).

All the galaxies in the HRS sample have been observed by the Two Micron All Sky Survey (2MASS; \citealp{jarrett2003}) in the J (1.25 $\mu$m), H (1.65 $\mu$m) and K$_{S}$ (2.15 $\mu$m) bands. We adopt the $B-V$ colour-dependent stellar mass-to-light ratio relation based on the H band luminosity
\begin{eqnarray}\label{eq:stellarmass}
\log (M_{*}/{\rm M_{\odot}}) = -0.059 + 0.21(B-V) + \log (L_{H}/{\rm L_{\odot}})
\end{eqnarray}
from \cite{bell2003}, which assumes a \citet{salpeter1955} IMF. B and V band observations are available from the GOLDMine database (\citealp{gavazzi2003}) for 182 of the 260 galaxies and morphologically averaged $B-V$ colours are used when optical observations are unavailable. Morphologies are taken from the {\it NASA/IPAC Extragalactic Database} (NED). Bell et al. (2003) report the errors in the mass-to-light ratio increase with $B-V$ colour, typically between 0.1 and 0.2 dex for galaxies with blue optical colors. Including some error associated with the choice of IMF, our typical errors in stellar mass are $<$0.3 dex.

Observations from the GALEX \citep{martin2005} GR2 to GR4 data releases in the near-ultraviolet (NUV; $\lambda$=2316 \AA: $\Delta \lambda$=1069 \AA) band were available for all but one galaxy. NUV magnitudes were obtained by integrating the flux over the galaxy optical size, determined at the 
surface brightness of $\mu$(B) = 25 mag arcsec$^{-2}$, with a typical uncertainty in the NUV photometry of $\sim$0.1 mag. We use the $L_{TIR}$/$L_{UV}$ method and the age-dependent relations of \citet{cortese2008} to correct the UV magnitudes for internal dust attenuation. The TIR luminosity is obtained from IRAS 60 and 100 $\mu$m fluxes or, in the few cases when IRAS observations are not available, using the empirical recipes described in \citet{cortese2006a}. The typical uncertainty in the NUV dust attenuation is $\sim$ 0.5 mag. Using the corrected\clearpage 

\begin{figure*}[!hp]
\begin{center}
\includegraphics[width=0.29\textwidth]{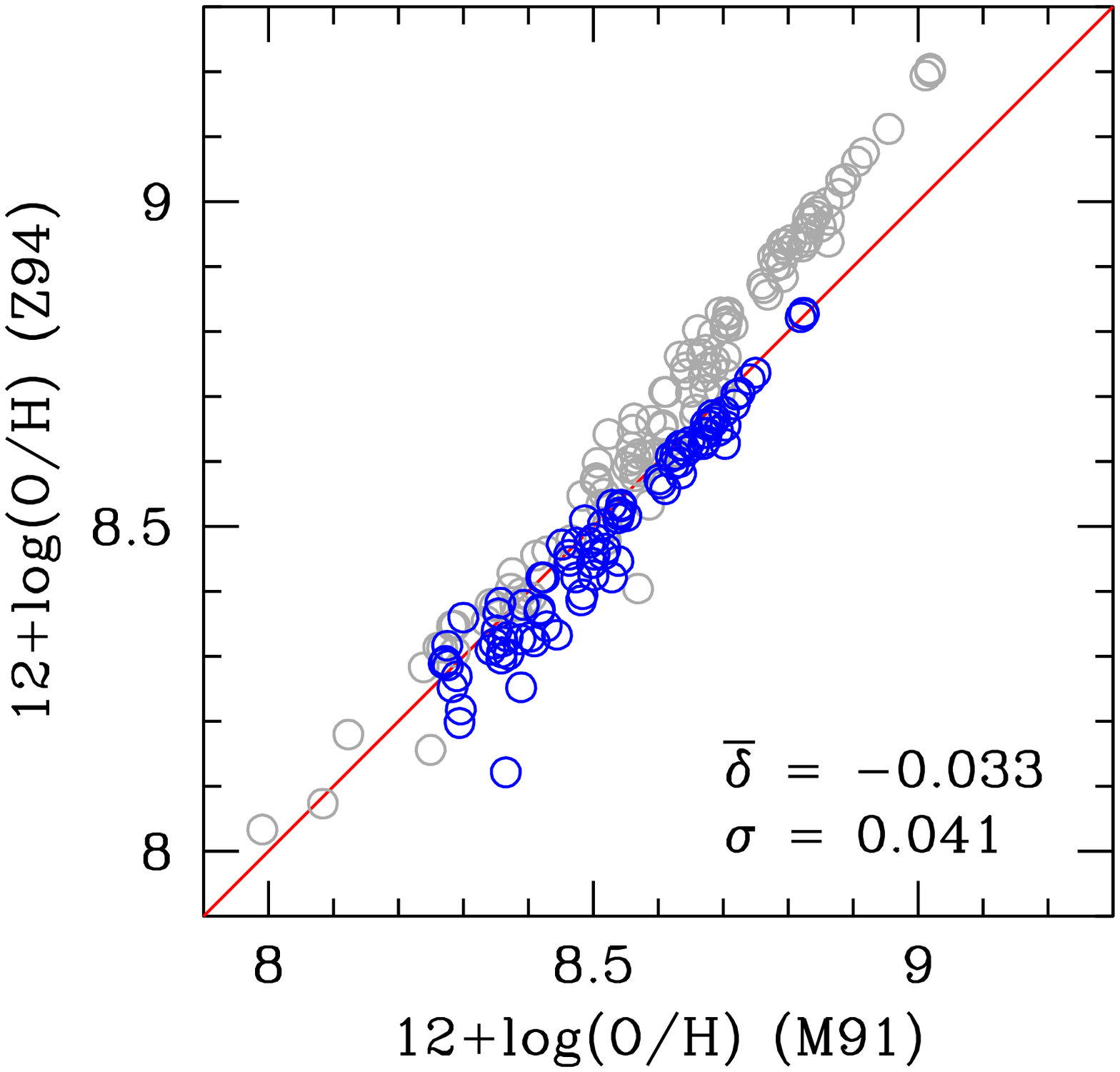}
\includegraphics[width=0.29\textwidth]{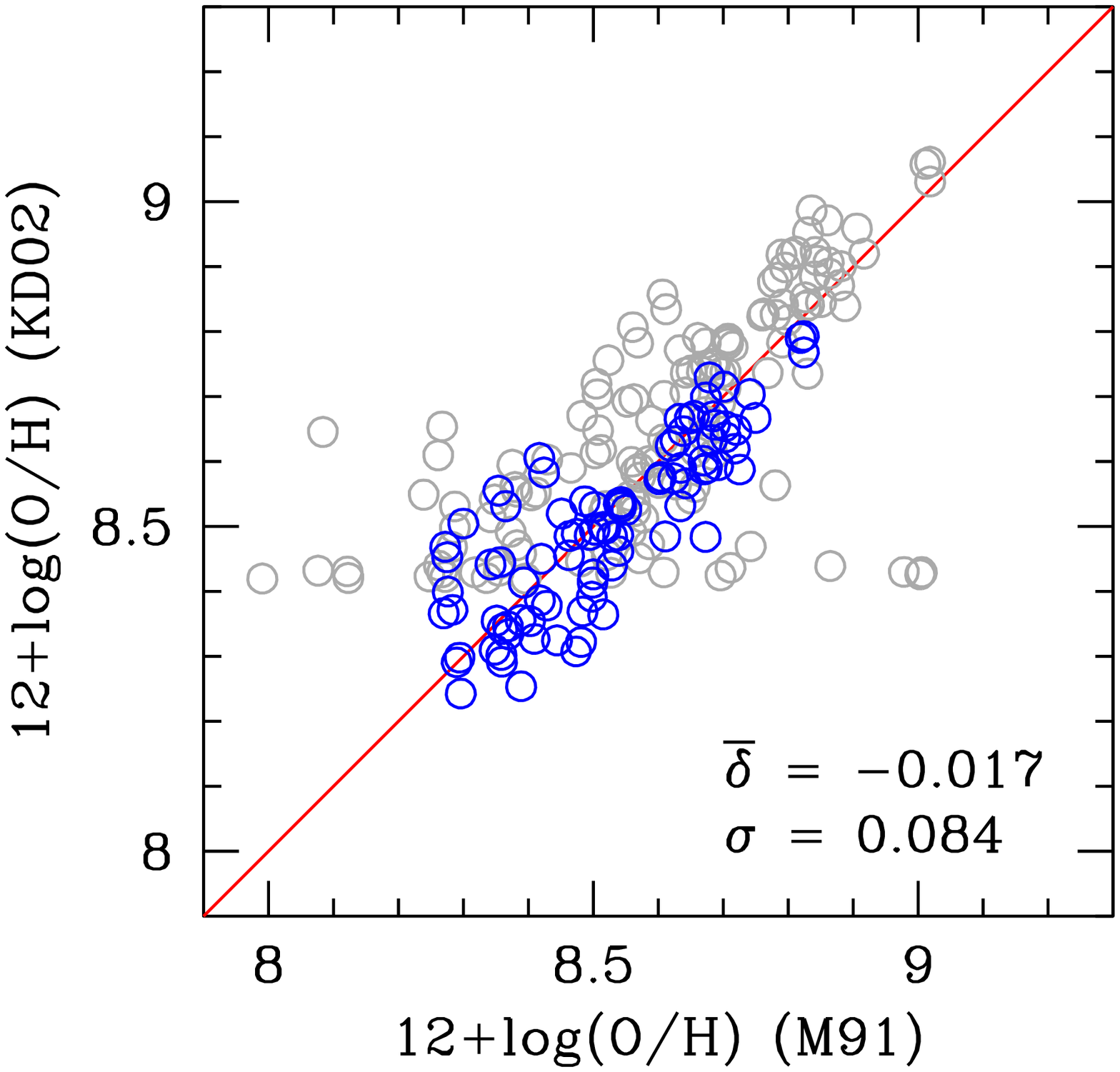}
\includegraphics[width=0.29\textwidth]{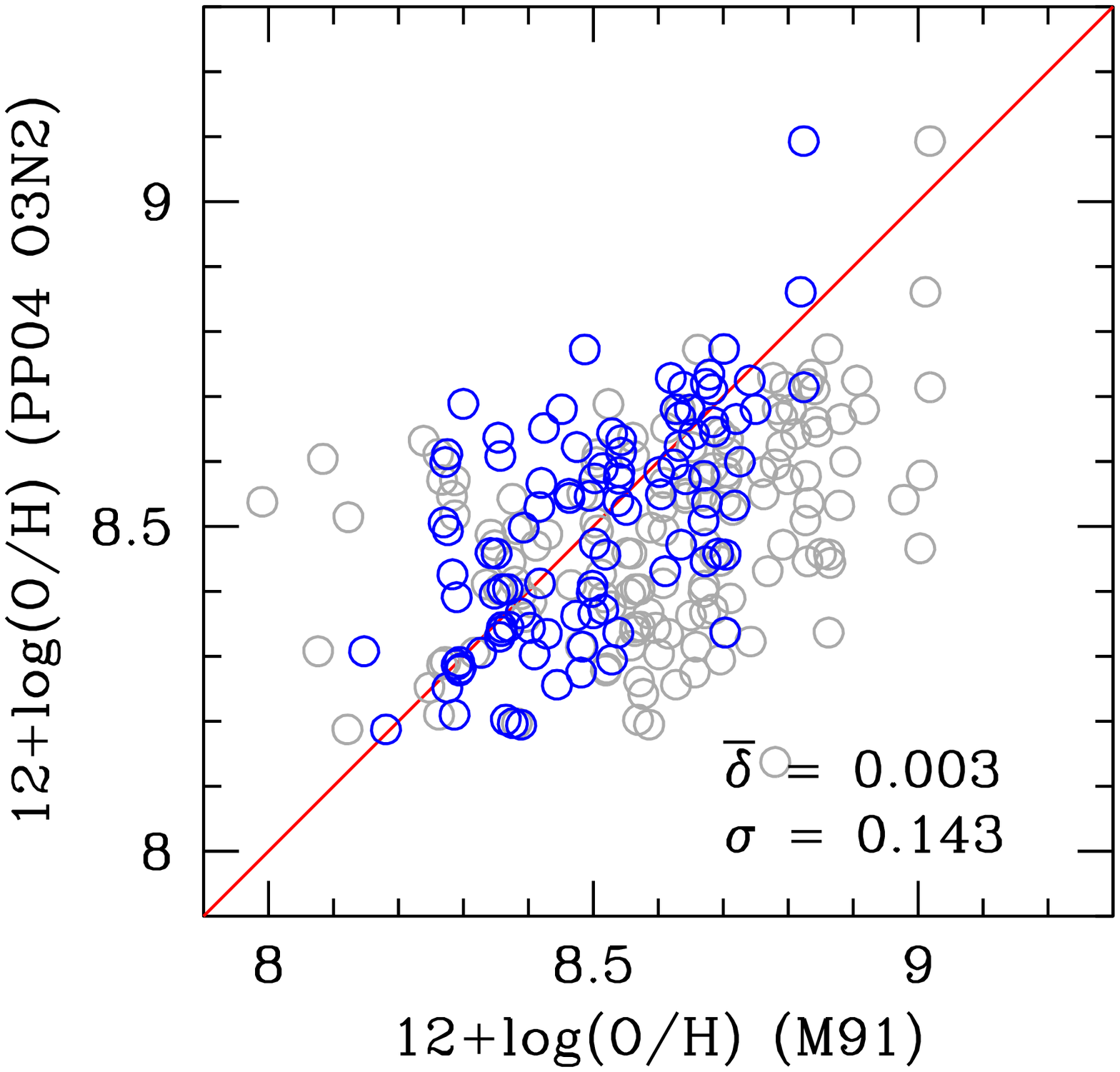}
\includegraphics[width=0.29\textwidth]{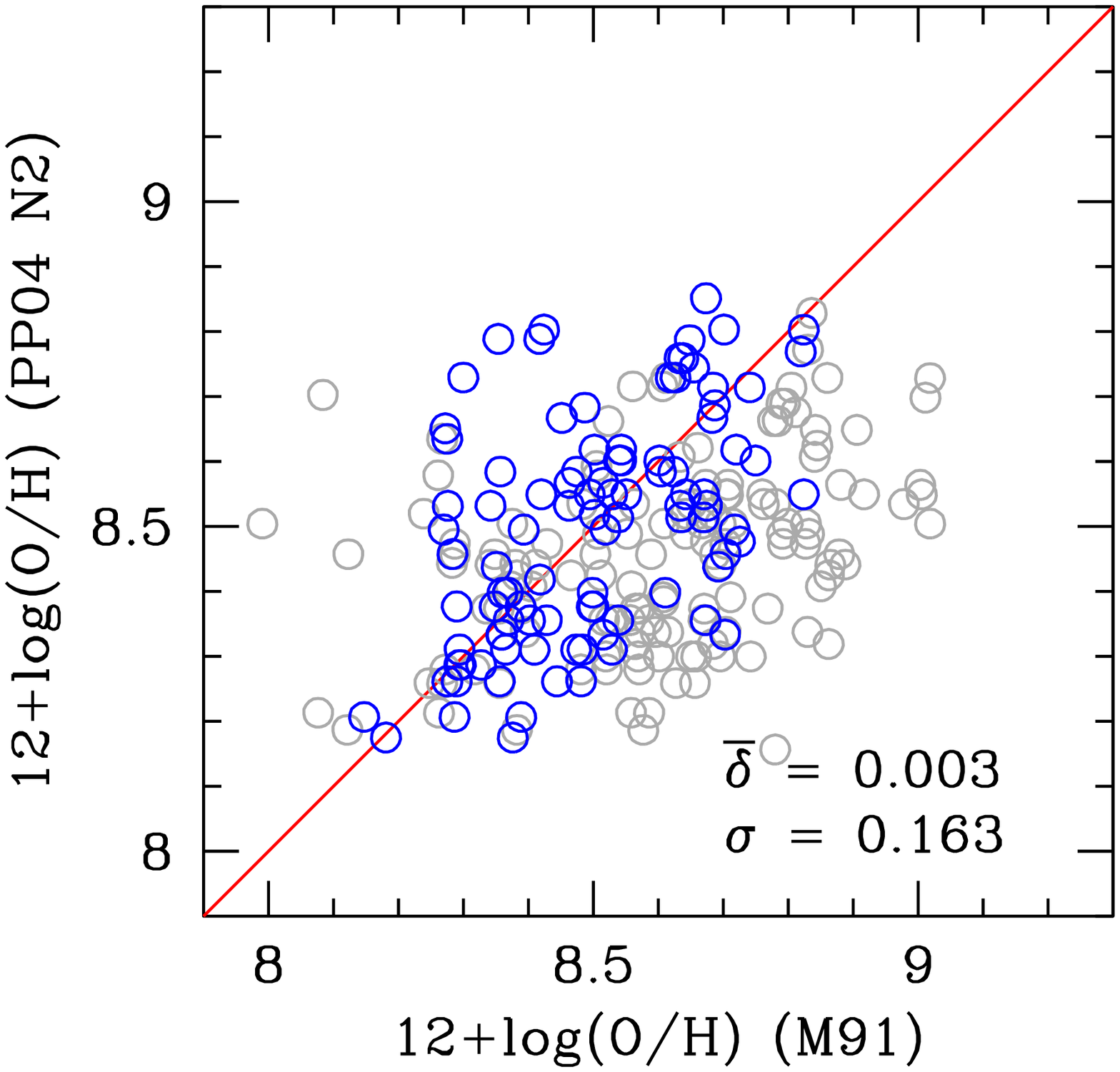}
\includegraphics[width=0.29\textwidth]{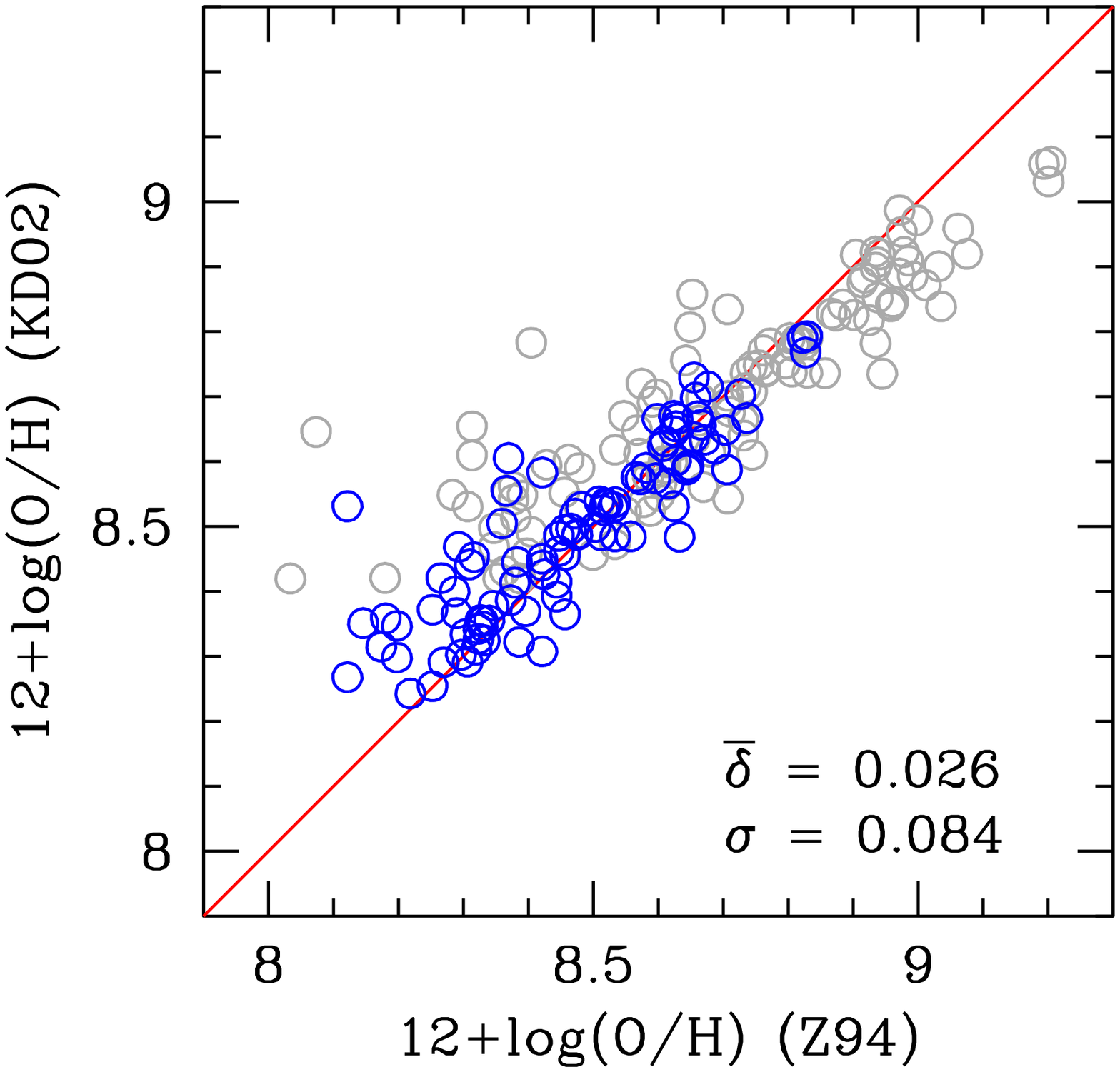}
\includegraphics[width=0.29\textwidth]{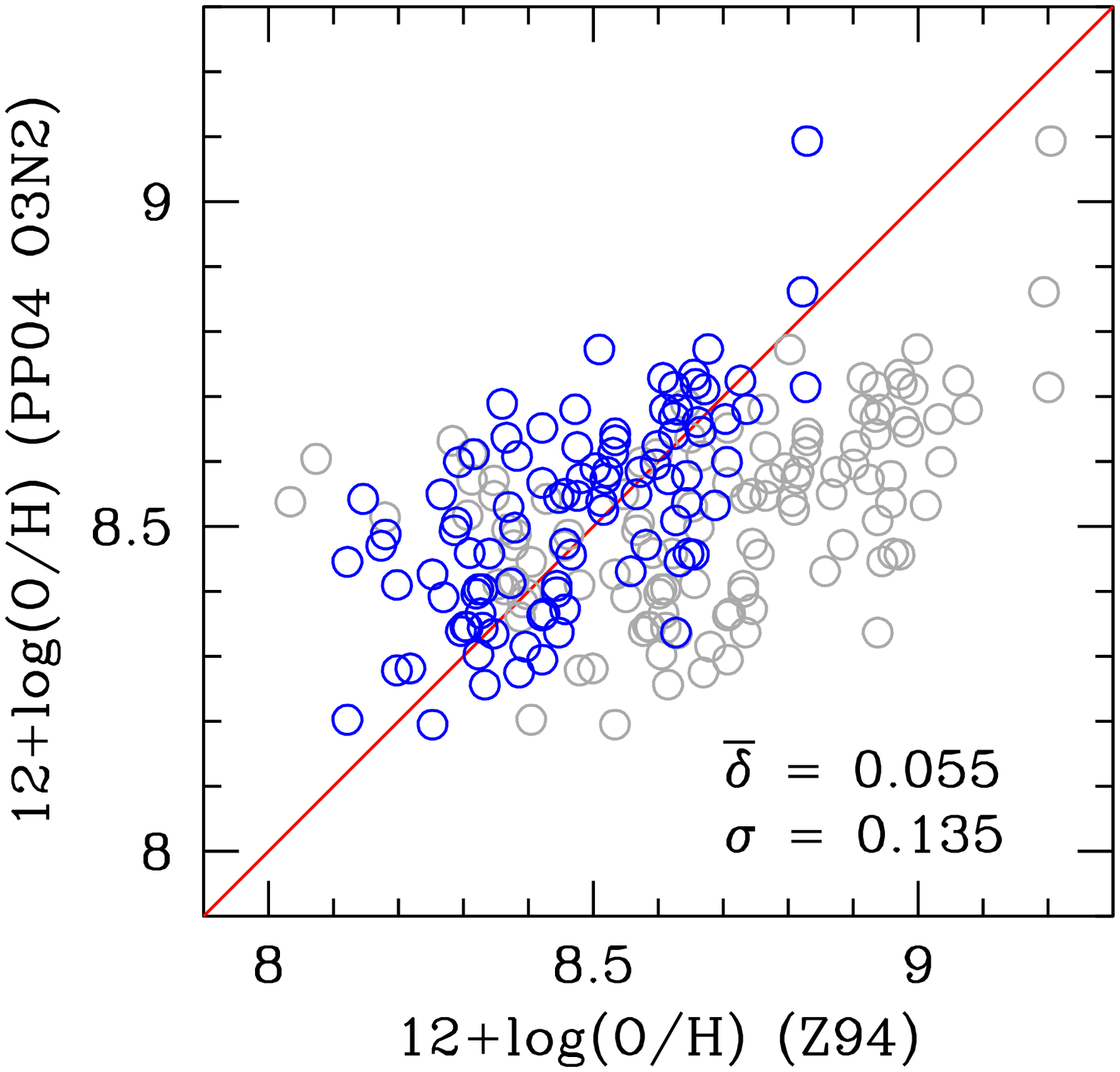}
\includegraphics[width=0.29\textwidth]{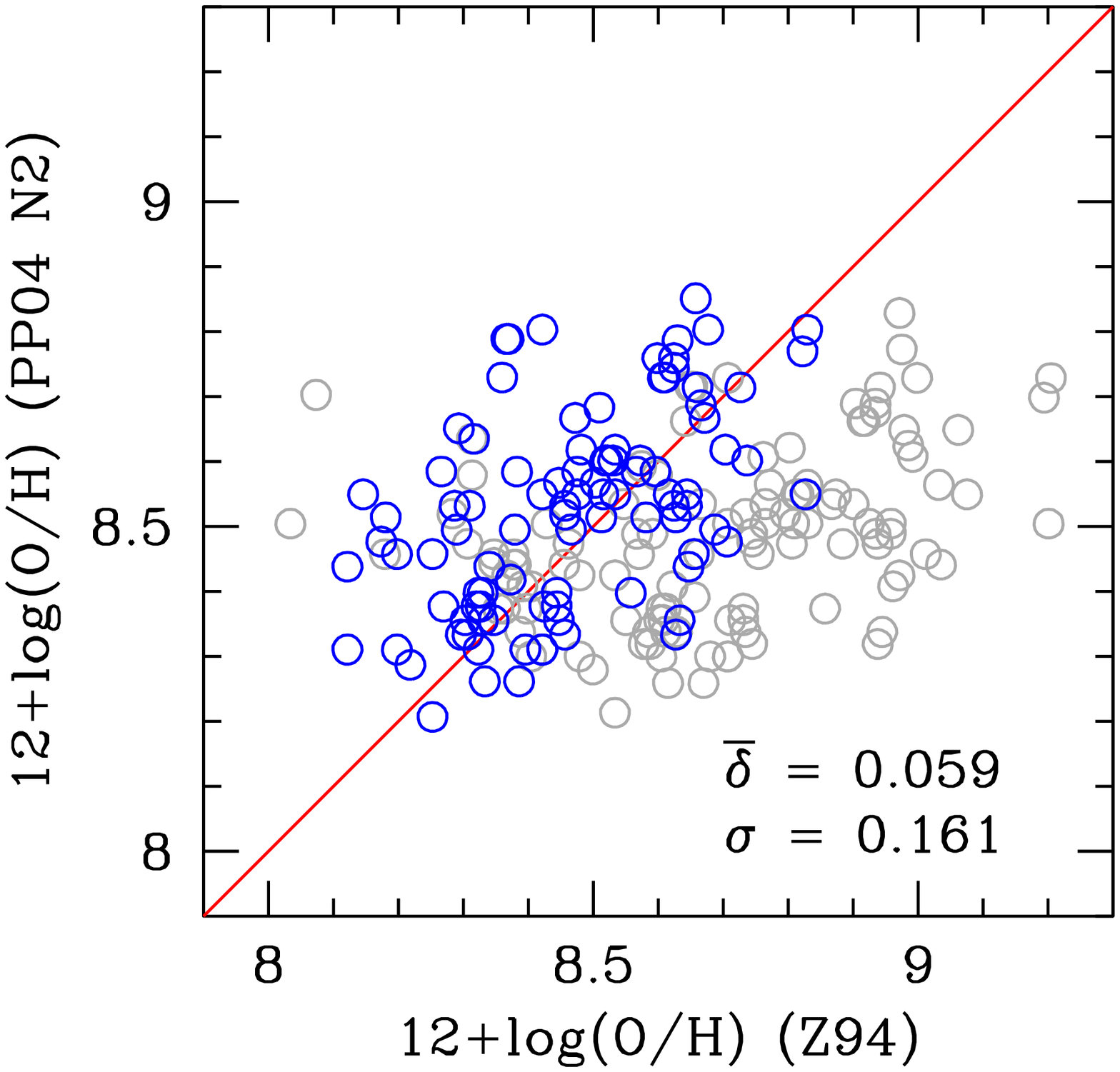}
\includegraphics[width=0.29\textwidth]{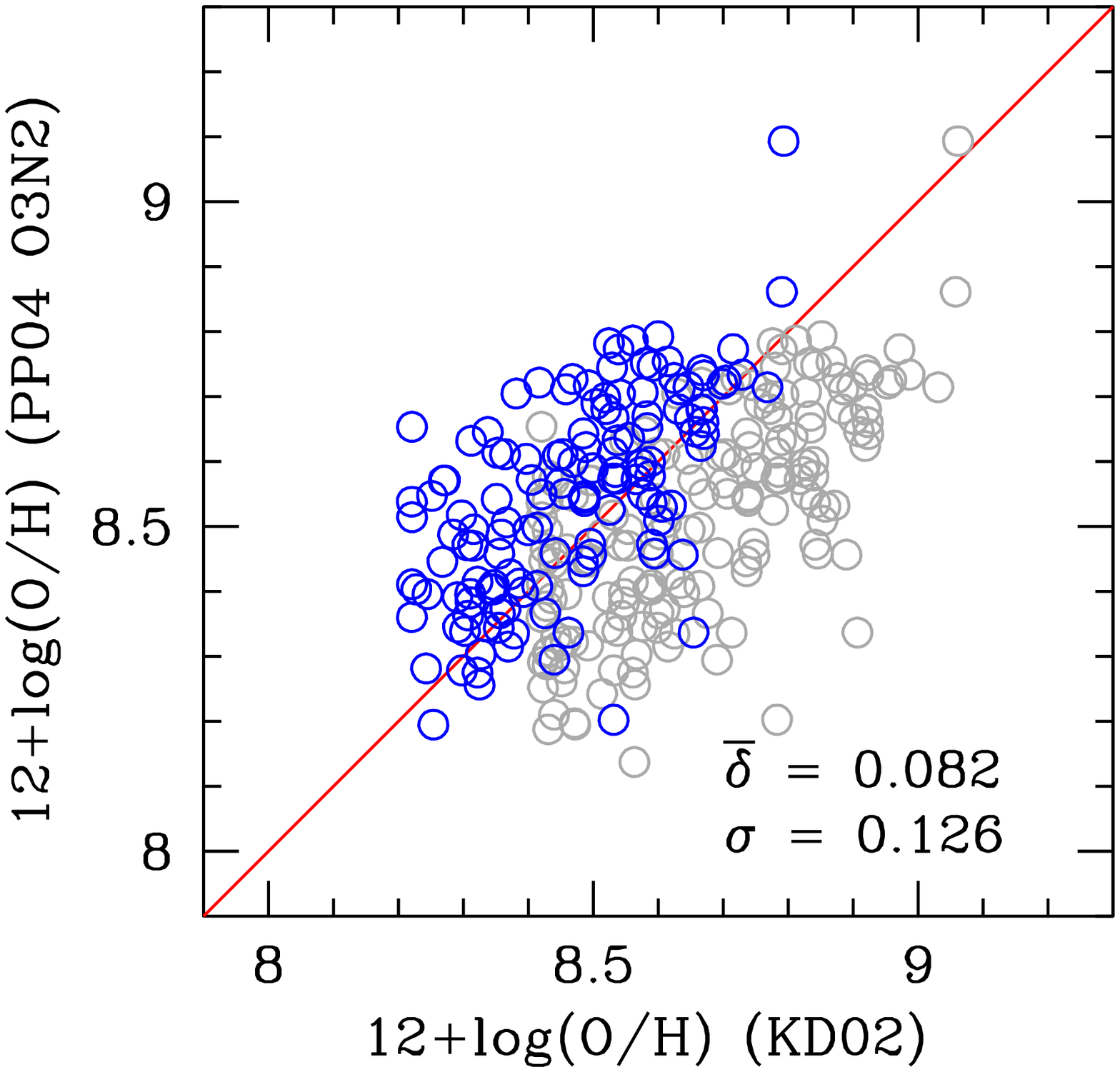}
\includegraphics[width=0.29\textwidth]{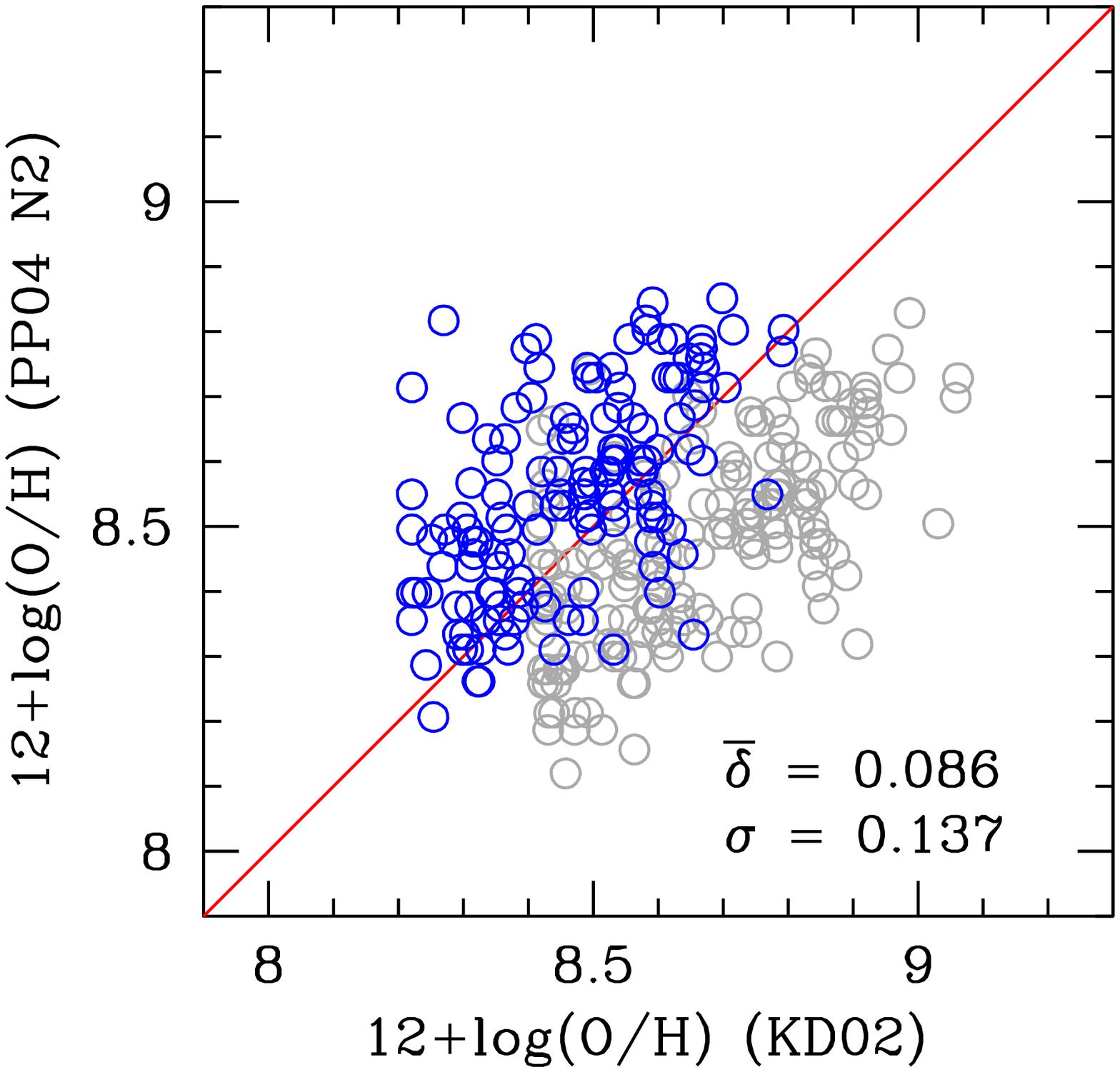}
\includegraphics[width=0.29\textwidth]{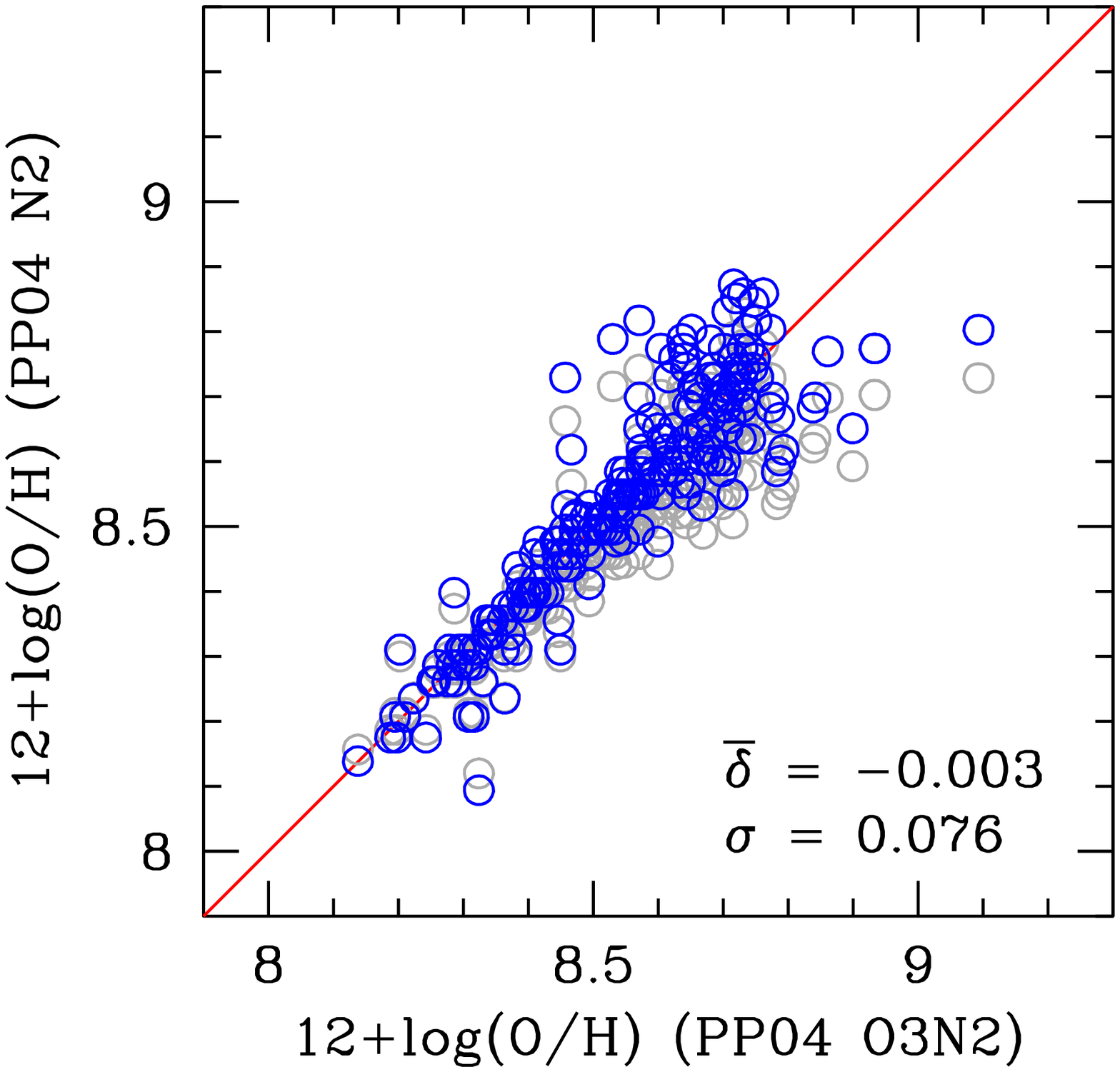}
\end{center}
\caption[Z-Z relationships with PP04 O3N2 as the base metallicity]{The relationships between the metallicities obtained from the five calibration methods once converted into the PP04 O3N2 base metallicity (blue open circles). The 1:1 line is shown in red and the unconverted metallicities from each calibration are shown for comparison (grey open circles). The scatter, $\sigma$, is the standard deviation of the residuals, $\delta$, between the two plotted metallicities converted into the PP04 O3N2 base metallicity, and $\bar{\delta}$ is the mean of these residuals.}\label{fig:o3n2}
\end{figure*}
\clearpage

\noindent
NUV measurements, star formation rates (SFR) are estimated from the conversion relations of \citet{iglesias2006}, who find that the relation
\begin{eqnarray}\label{eq:sfrs}
\log SFR_{NUV,cor} \, ({\rm M_{\odot} \, yr^{-1}}) = \log L_{NUV,cor} \, ({\rm L_{\odot}}) - 9.33
\end{eqnarray}
provides the rate to an accuracy $<$20\% under the assumption of a \citet{salpeter1955} IMF. 

Single-dish \hi~21 cm line emission data, necessary for quantifying the \hi~content of galaxies, were taken from \cite{gavazzi2003}, \cite{springob2005}, \cite{giovanelli2007}, \cite{kent2008} and NED. Estimates of atomic hydrogen mass or upper limits are available for $\sim$97\% (252/260) of the late-type galaxies. We estimate the \hi~deficiency parameter, $DEF(HI)$, as defined by \cite{haynesgio1984}: i.e., 
the difference, in logarithmic units, between the observed \hi~mass and the value expected 
from an isolated galaxy with the same morphological type $T$ and optical linear diameter $D$:
\begin{eqnarray}\label{eq:defhi}
DEF(HI) = \log M^{exp}_{HI}(T,D) - \log M^{obs}_{HI}.
\end{eqnarray}

\begin{figure}[!t]
\begin{center}
\includegraphics[width=0.35\textwidth]{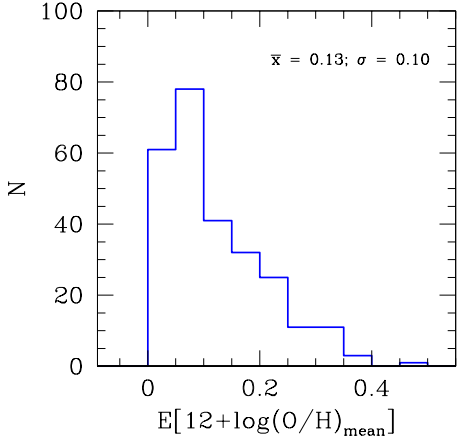}
\end{center}
\vspace{-0.75 cm}
\caption[The distribution of the errors in the final metallicity values]{The distribution of the errors in the final metallicity values based on the PP04 O3N2 calibration.}\label{fig:finalerrors}
\end{figure}

We use the equation in \cite{haynesgio1984} to calculate the 
expected \hi~mass from the optical diameter. The four revised coefficients that vary with morphological type are taken from \cite{solanes1996} for Sa-Sab, Sb, Sbc, Sc types, and from \citet*{boselligavazzi2009} for Scd to Irr types.
Given its large uncertainty, in the following, we will mainly use the \hi~deficiency to select those 
galaxies which have likely lost a significant 
amount of atomic hydrogen. In detail, we use a threshold of $DEF(HI)=$ 0.5 to discriminate 
between `\hi-deficient' and `\hi-normal' galaxies. \hi-deficient systems are thus objects with $\geq$70\% less atomic hydrogen than expected for isolated objects of the same optical size and morphological type. Out of all 260 HRS late-types, 33\%(64\%) are \hi-deficient(-normal), with 71\% of the \hi-deficient late-types residing in the Virgo cluster.

\begin{center}
\begin{table*}[t]\renewcommand{\arraystretch}{1.4}
\begin{center}\caption[A summary of the different calibrations used to estimate the gas-phase metallicity]{A summary of the calibrations used to estimate the gas-phase metallicity.}\label{tab:zcalsummary}
\tiny \begin{tabular}{l l c c  }
\cline{1-4}
Reference & Method & Emission Lines &  Limits  \\
\cline{1-4}\renewcommand{\arraystretch}{1.2}
\citealp{m91} & M91 & [\oii], \hb , [\oiii] &  7.1 $<$ \zzz $<$ 9 \\
\citealp{z94} & Z94 & [\oii], \hb , [\oiii] &  \zzz $>$8.4 \\
\citealp{kd02} & KD02 & [\oii], [\nii] &  \zzz $>$8.4\\
\citealp{pp04} & PP04 N2 & [\nii], \ha &  7.17 $<$ \zzz $<$ 8.87 \\
\citealp{pp04} & PP04 O3N2 & \hb , [\oiii], [\nii], \ha &  8.12 $<$ \zzz $<$ 9.05 \\
\cline{1-4}
\end{tabular}
\end{center}
\end{table*}
\end{center}

\section{Estimating metallicity}\label{sec:metestimates}

In this section, we discuss in detail the new method adopted for estimating global oxygen abundances. In order to test the accuracy of the method and examine the relationships between metallicities derived from different calibrations, we include, in addition to the HRS sample, \textit{all} available spectroscopic observations for galaxies in the GOLDMine database \citep{gavazzi2003}. The observations presented in \citet{gavazzi2004} cover galaxies in the Virgo cluster and other nearby clusters - Coma and A1367, Cancer, A262 and Centaurus - plus 10 isolated objects, thereby increasing the number of observed galaxies to 478 objects. We stress the use of the 478 observations is restricted to testing our methodology in this section, whereas we use the HRS sample of 260 late-types for the main analysis.

Oxygen abundances are estimated from measurements of emission line fluxes using five different calibrations from the literature. In order to compare the metallicity estimates given by the different metallicity indicators and hence determine the most accurate calibration, we use calibrations derived from both empirical observations and theoretical models. The two calibrations presented in \citet[hereafter PP04]{pp04} are empirical estimates of the oxygen abundances calibrated from observations of \hii regions, using the [\nii] $\mathrm{\lambda}$6584/\ha \ (N2) and [\oiii] $\mathrm{\lambda}$5007/[\nii] $\mathrm{\lambda}$6584 (O3N2) ratios. The calibrations of \citet[M91]{m91} and \citet[Z94]{z94} are theoretical methods based on the \rz \ $\equiv \log ([$\oii$] \lambda 3727+[$\oiii$] \lambda 4959+[$\oiii$] \lambda 5007) /$ \hb \ ratio. Finally, the \citet[KD02]{kd02} calibration uses the [\nii] $\mathrm{\lambda}$6584 / [\oii] $\mathrm{\lambda}$3727 ratio. All these calibrations do not require knowledge of the [\oiii] $\lambda$4363 emission line and are indirect measures of the metallicity. We note that the \rz \ ratio is double-valued; we use the [\nii]/[\oii] ratio to break the \rz \ degeneracy for the M91 and Z94 calibrations, as detailed in Appendix A of KE08. Table \ref{tab:zcalsummary} provides a summary of the different calibrations used in this study.

Since the optical emission lines are susceptible to contamination due to AGN emission, which may lead to incorrect estimates of the oxygen abundance, our first step is to identify and eliminate potential AGN-hosts and galaxies displaying AGN-like characteristics from our samples. The BPT (\citealp*{bpt81}) diagnostic diagram based on the [\nii]/\ha \ and [\oiii]/\hb \ line ratios is used to classify the spectra of galaxies (see Fig.~\ref{fig:bpt}). We remove all galaxies that are above the \citet{kauffmann2003} demarcation line between star-forming and AGN-host galaxies. Furthermore, objects for which a measurement of [\oiii]/\hb \ is not available (plotted at log([\oiii]/\hb) = -1.3 in Fig.~\ref{fig:bpt}) may host AGN or demonstrate AGN-like activity. We therefore also remove those objects with log([\nii] $\lambda$6584/\ha) $>$ -0.4, which are classified as LINER, Seyfert or composite/transition galaxies with the \citet{decarli2007} scheme. These conservative cuts remove 36 potential AGN-host galaxies, 23 of which are members of the HRS sample. Taking into account that the validity ranges of the various metallicity calibrations tend to exclude emission line ratios characteristic of AGN-hosts, these cuts remove only 6 HRS galaxies with useable metallicities that could contaminate the analysis.  

We then apply each of the calibrations to the emission line flux measurements for each galaxy. As previously mentioned, the spectra are corrected for internal and galactic extinction, and the \ha \ and \hb \ lines are corrected for underlying stellar absorption (see Boselli et al. 2012). Systems lacking a measure of C(\hb) due to undetected \hb \ emission only have metallicities estimated with the PP04 N2 calibration. Unlike the other emission line ratios exploited by the remaining calibrations, the [\nii] and \ha \ lines used in this calibration have very close wavelengths and are similarly affected by dust extinction. Yet, because the values of the reddening function adopted for extinction corrections are similar for the [\nii] and \ha \ lines, the ratio does not require a measurement of C(\hb) and therefore can still be used for estimating oxygen abundances.

We first examine the relationships between the estimates derived from the different calibrations. A visual inspection of metallicity-metallicity diagrams, where each calibration method is plotted against the other four methods (Fig. \ref{fig:o3n2}, grey circles), confirms the significant discrepancies which arise when using different calibration methods which have \textit{not} been converted into a `base' metallicity, as discussed in KE08. The mean residual metallicity given by $\bar{\delta}$, where $\delta = x - y$, is 0.17 dex. The standard deviation of the difference between the estimates of the two calibrations is lowest for the PP04 O3N2 and N2 calibrations ($\sigma =$ 0.07 dex) and highest for between Z94 and PP04 N2 ($\sigma =$ 0.25 dex). It is evident that the estimates, and thus the discrepancies and the shape of the relations between different calibrations, are consistent with those found in previous studies and in KE08. This is encouraging when one considers that the two datasets were obtained in different ways: KE08 use spectroscopy from SDSS fibers, whereas this work uses integrated, drift-scan spectroscopy. 

The situation improves once the systematic discrepancies between the different calibrations are removed by converting the results into a base metallicity, such that all the abundance estimates from different calibrations are comparable. By fitting the relationship between the various results of the calibrations, KE08 were able to eliminate the systematic discrepancies and produce comparable metallicity measurements from different calibration methods. The conversion relations enabling a metallicity from one method to be converted into any other metallicity follow $y = a + bx + cx^{2}+dx^{3}$ where $y$ is the final or `base' metallicity in \zzz \ units, $x$ is the original metallicity from an alternate calibration and $a$-$d$ are the third order coefficients, as presented in Table 3 of KE08. We convert each metallicity estimate from the five calibrations into the PP04 O3N2 calibration, because it results in very small residual metallicity discrepancies when used as a base calibration (KE08, Fig 4). Fig. \ref{fig:o3n2} also presents metallicity-metallicity diagrams for each calibration plotted against the other four calibration methods after this conversion into the PP04 O3N2 base metallicity. It is evident that the systematic discrepancies between the five calibrations are significantly reduced and remain consistent with KE08. The mean residual metallicity is $\bar{\delta}=$0.03 dex and the average standard deviation of the differences between the estimates of the calibrations is $\sigma =$0.11 dex.

The final metallicity estimates of each galaxy are derived from the mean of all the applicable calibrations (i.e. upto five metallicity estimates available per galaxy) converted into the PP04 O3N2 base metallicity and weighted by the errors associated with each calibration method. These calibration errors are obtained from the standard deviation of the scatter when comparing the metallicities from a particular calibration with the those of the remaining four calibrations. The errors in the M91, Z94, KD02, PP04 O3N2 and N2 are 0.12, 0.11, 0.12, 0.10 and 0.10 dex, respectively. The overall error is calculated from the error-weighted normalised mean metallicity to give the standard weighted error. Fig. \ref{fig:finalerrors} shows the distribution of the errors in our final metallicity measurements. The average error is 0.13 dex. We obtain oxygen abundances for 272 galaxies from 442 observations\footnote{Even after excluding potential AGN-host galaxies, not all spectroscopic observations yield a metallicity estimate, due to either the absence of the required emission lines or emission line ratios which fall outside of the validity ranges of the calibration methods.}, presented in Table 2, of which 169 galaxies are HRS late-types; 75 (94) galaxies are Virgo cluster (non-Virgo, field) objects. Of all the HRS galaxies with oxygen abundances, 36\% use 4 or 5 calibrations, 51\% use 2 or 3 calibrations and only 13\% of the sample rely on one calibration. Thus, it is now possible to explore the nature of the M-Z relation in different environments using the HRS sample of galaxies combined with these new estimates of the gas-phase metallicities.

\section{Defining the M-Z relation}

\begin{figure*}[ht]
\begin{center}
\includegraphics[width=0.95\textwidth]{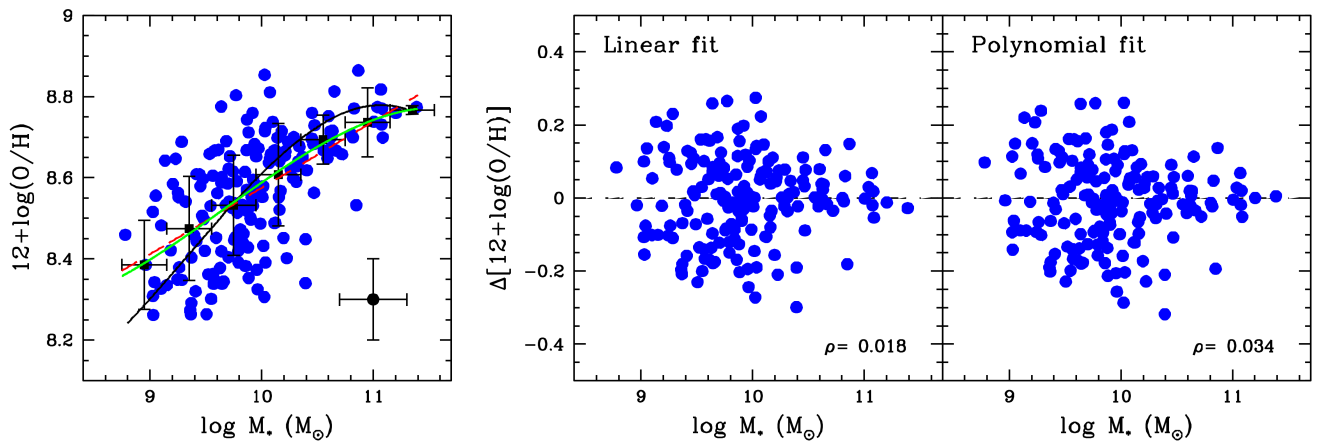}
\end{center}\vspace{-0.5cm}
\caption[The mass-metallicity relationship for the HRS+ sample, highlighting estimates derived from single measurements]{\textit{Left panel}: The relationship between stellar mass and metallicity for the HRS sample of galaxies (blue circles), using metallicity estimates based on the PP04 O3N2 base calibration. The KE08 best fit (solid black line) and our linear and polynomial fits (dashed red and solid green lines, respectively) to the binned data (black squares) are presented. Error bars show the 1-$\sigma$ scatter in each bin. \textit{Right panels}: The residual metallicities as obtained from the linear and polynomial best fits.}\label{fig:mzrshighlighted}
\end{figure*}

First, we investigate the relationship between stellar mass and gas-phase oxygen abundance for the HRS sample, presented in Fig. \ref{fig:mzrshighlighted}. A positive correlation exists between the two quantities with a Spearman coefficient of rank correlation of $\rho = 0.61$, corresponding to a probability P($\rho$) $>$ 99.9\% that the two variables are correlated.
Also displayed in Fig. \ref{fig:mzrshighlighted} is the best fit to the PP04 O3N2-based M-Z relation presented in KE08 (renormalised to a Salpeter IMF following \citealp{bell2003}), which is fairly consistent with our own observations. To obtain our best fit M-Z relation, we use the IDL polyfit task to perform a least squares fit to the mean metallicity in bins of fixed stellar mass. We find that a linear fit would most likely suffice to describe the M-Z relation for this sample. However, recent studies have reported that the M-Z relation flattens with increasing stellar mass and that the relation is best described using a polynomial function (see e.g. KE08 and references therein). The use of a polynomial has a physical as well as observational basis, since simple chemical evolution models predict that as gas is consumed via star formation, with a consequent increase of the stellar mass, the metallicity reaches a maximum upper limit that is determined by the mass of metals freshly produced in stars and ejected into the ISM, i.e. the yield (see e.g. \citealp{edmunds1990}; \citealp{erb2008}). Hence, we also fit a polynomial function to the binned data, obtaining 
\begin{eqnarray}
y = 22.800 - 4.821 x + 0.519 x^{2} - 0.018 x^{3}
\end{eqnarray}
where $x$ is the logarithm of the stellar mass in units of M$_{\odot}$ and $y$ is the oxygen abundance in \zzz \ units. This best fit M-Z relation is fairly consistent with those presented in KE08, with discrepancies between the fit presented here and the KE08 fits being within the errors over most of the range in stellar mass. The disagreement in the fit arises at stellar masses lower than $\sim$10$^{9}$ M$_{\odot}$, due to the HRS sample having too few low mass galaxies to accurately constrain the y-intercept of the M-Z relation, compared to the larger range of the sample used in KE08. The metallicity residuals from the linear and polynomial fits are presented in Fig. \ref{fig:mzrshighlighted}. This figure is important since the\clearpage

\noindent
residuals represent the scatter in the M-Z relationship, and it is important that the fitted relationship does not introduce any artificial trends across the mass range, in order to accurately study the origin of the scatter. Both residual distributions are remarkably similar, with no trends introduced from the fits. Indeed, the Spearman correlation coefficients for each of the panels are both much smaller than 0.04, where a coefficient of zero indicates no observed trend in the data. The scatter in the M-Z relation is 0.11 dex, which is only slightly larger than the scatter reported in T04 and comparable to the 0.1 dex error associated with the metallicity estimates. Therefore, the choice of linear or polynomial fit does not make a significant difference to the distribution of the residual metallicities, yet we proceed with the analysis using the residuals from the polynomial fit for the physical reasons expressed above.

\section{Gas content and SFR}

In the previous section, a correlation between the stellar mass and gas-phase oxygen abundance was confirmed. The most interesting observation is, as previously noted by T04, the tightness of the correlation, as the scatter from the fitted M-Z relations are only $\sim$0.1 dex, which is comparable to the error associated with the metallicity estimates. This observation is quite remarkable when consideration is given to the plethora of processes that potentially affect both quantities. In this and the following section, we will investigate the origin of the scatter in the M-Z relation, by considering properties that could, in theory, influence the link between stellar mass and metallicity. Two natural quantities that have been proposed in the past are the gas fraction and star formation rate. 

A link between gas content and metallicity is supported on theoretical grounds. It is well known that a solution to a simple closed-box model for chemical evolution relates gas content to metal content (see \citealp{edmunds1990} and references therein). Recently, \citet{zhang2009} investigated the role of gas fraction on the M-Z relation by using a sample of $\sim$10$^{5}$ SDSS galaxies. They showed that galaxies with a lower gas content often display enhanced metallicities at a fixed stellar mass and found a systematic change in the gas fraction along the M-Z relation. However, it is important to note that \citeauthor{zhang2009} did not use \hi \ masses obtained from 21 cm observations, but empirical ones estimated from a recipe based on colour and stellar mass surface density. Using the HRS sample, we are now able to investigate the importance of the gas fraction using direct measurements of the \hi \ mass, instead of indirect estimates. Following previous work, we define the gas fraction as the ratio of the gas mass to the sum of the gas and stellar mass,
\begin{eqnarray}\label{eq:mu}
\mu = \frac{M_{gas}}{M_{*} + M_{gas}}
\end{eqnarray}
such that it represents the amount of gas which has not yet been turned into stars. The gas mass is considered to be the mass of \hi \ ($M_{HI}$) with a correction for neutral helium
$M_{gas}$ = 1.32 $M_{HI}$. We note that this is a `hybrid' gas fraction, since it does not take into account the contributions of molecular hydrogen. The H$_{2}$ content comprises at least 20\% of the total gas mass and should be given consideration. However, we currently lack a homogeneous dataset of CO maps for the HRS sample. There is also a large uncertainty of the CO-to-H$_{2}$ conversion factor, which varies with metallicity and other physical properties of the ISM in galaxies \citep{boselli2002}. For these reasons, we use only the homogeneous \hi \ data available for the HRS.  

\begin{figure}[t]
\begin{center}
\includegraphics[width=0.45\textwidth]{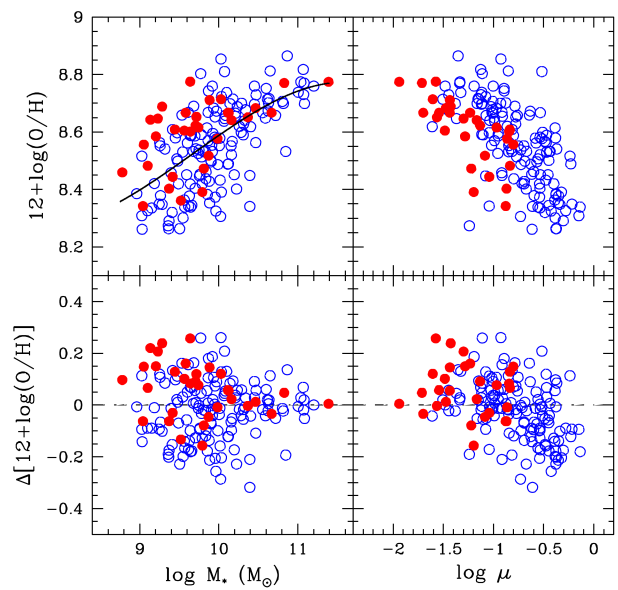}
\end{center}\vspace{-0.25cm}
\caption[The relationships between metallicity and gas fraction]{The relationships between stellar mass (\textit{left panels}) and gas fraction (\textit{right panels}) with oxygen abundance (\textit{upper panels}) and residual oxygen abundance (\textit{lower panels}) obtained from the best polynomial fit (solid black line in the upper left panel). Dividing the sample by \hi \ content demonstrates that gas content is anti-correlated with metallicity for \hi \ normal galaxies (blue open circles). Furthermore \hi \ deficient systems (red solid circles) are preferentially found at higher metallicities. Such trends remain when eliminating the stellar mass dependency.}\label{fig:gasfrachidef}
\end{figure}

\begin{figure}[!t]
\begin{center}
\includegraphics[width=0.4\textwidth]{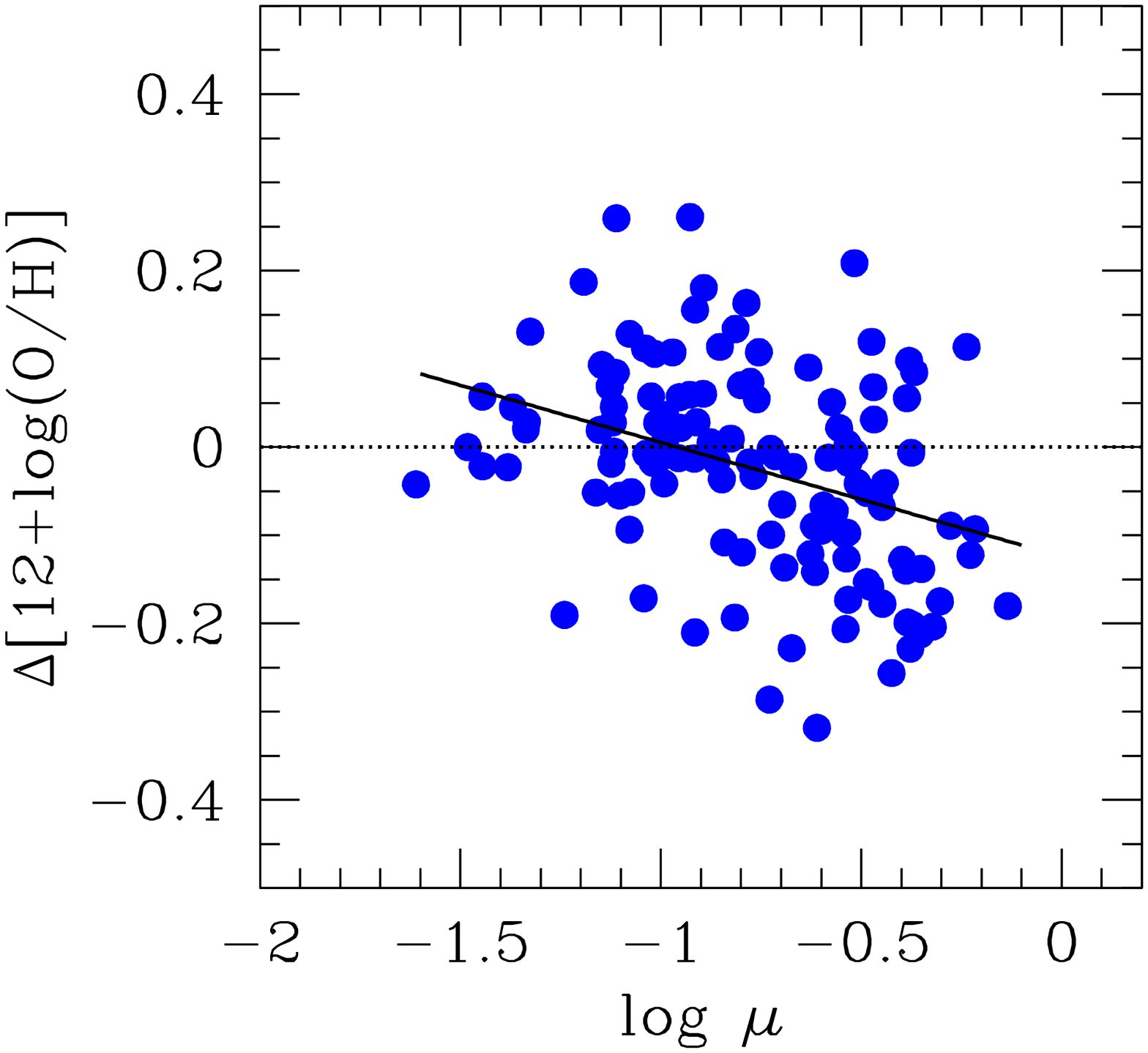}
\includegraphics[width=0.4\textwidth]{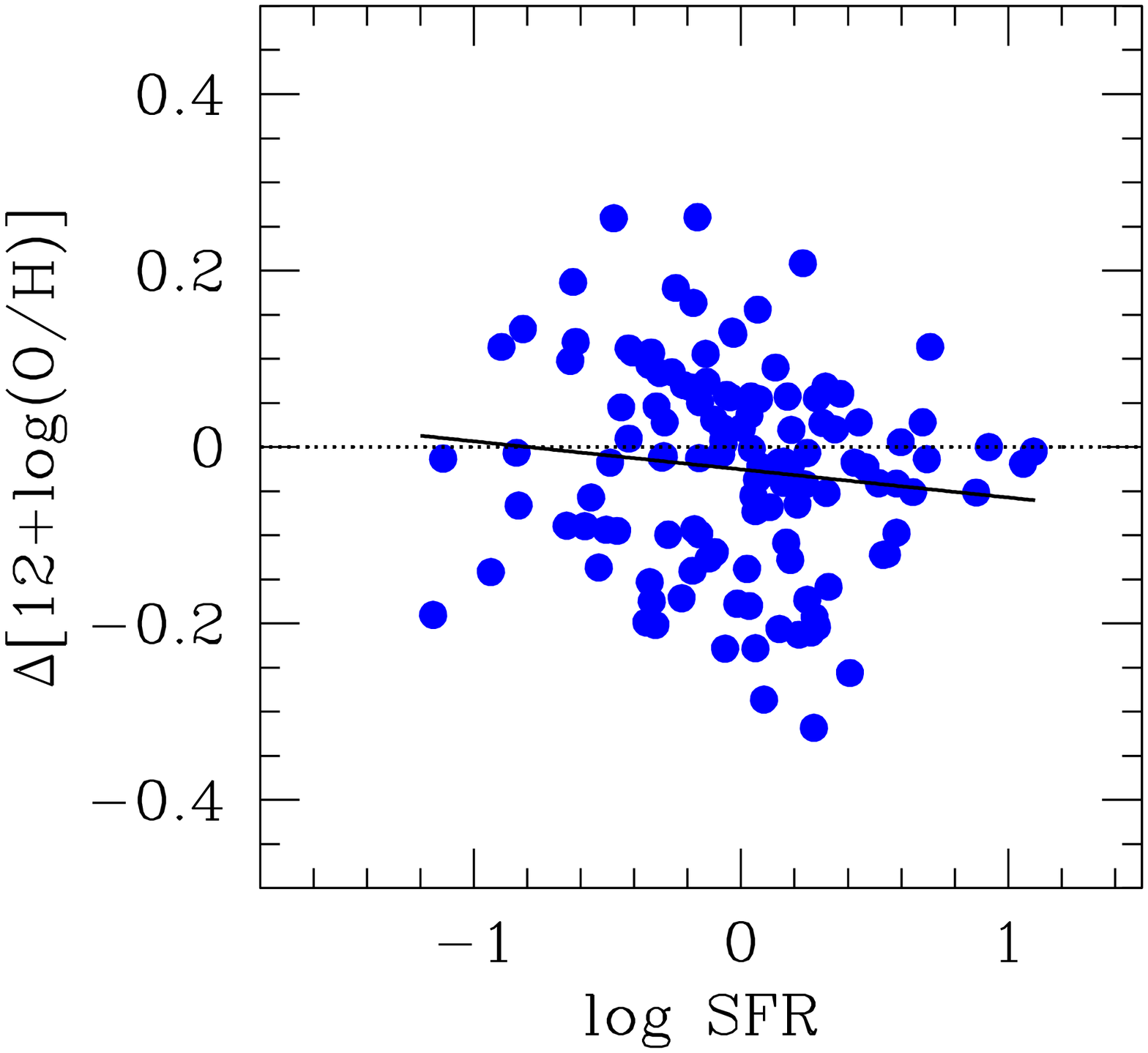}
\end{center}\vspace{-0.75cm}
\caption[Correcting the mass-metallicity relation]{\textit{Upper panel}: The best fit relationship between residual metallicity and the gas fraction, which is then used to reduce the scatter in the M-Z relation from 0.11 to 0.09 dex. \textit{Lower panel}: The same exercise is performed using the best fit relationship between residual metallicity and SFR, but the scatter in the M-Z relation is increased by 0.03 dex.} \label{fig:corrmetalsgf}
\end{figure}

Fig. \ref{fig:gasfrachidef} shows the relationship between the oxygen abundance and gas fraction. In the left panels of the diagram, \hi \ deficient galaxies highlighted on the M-Z relation and the residual M-Z relation (determined from the best fitting polynomial from the previous section) typically display higher metallicities than \hi \ normal galaxies;  \hi \ deficient galaxies are on average more metal-rich than their \hi \ normal counterparts by approximately 0.1-0.2 dex (we will return to this point in the next section). This immediately hints at a trend between gas content and metallicity. In fact, there is a strong trend between the oxygen abundance and the gas fraction of \hi \ normal galaxies: lower gas fractions tend to indicate a higher oxygen abundance. The correlation has a Spearman rank correlation of $\rho = -0.63$,  corresponding to a probability of correlation of P($\rho$) $>$ 99.9\%, and a scatter of 0.14 dex in the linear best fit relation. This trend between oxygen abundance and the gas fraction remains once the mass dependency is removed, with the relationship between residual oxygen abundance and gas fraction showing that the gas content may be responsible for the observed scatter in the M-Z relation. This correlation has a Spearman rank correlation of $\rho = -0.45$, corresponding to a probability of correlation of P($\rho$) $>$ 99\%, confirming theoretical expectations that at least part of the scatter in the M-Z relation is due to a different gas content at fixed stellar mass. Thus, using the relation between gas content and the residual metallicity for the \hi \ normal galaxies, we can investigate whether or not we can reduce the scatter in the M-Z relation. Using a linear relation between the gas fraction and residual oxygen abundance (Fig. \ref{fig:corrmetalsgf}) reduces the scatter around the original best fit M-Z relation from 0.11 dex to 0.09 dex. Such a reduction in scatter is a minor improvement with respect to the original relation. This might be due to the fact that the intrinsic error in the metallicity estimates does not allow us to reduce the scatter of the relation further. 

Two recent studies by \citet{laralopez2010} and \cite{mannucci2010} demonstrate that stellar mass and metallicity form a so-called fundamental plane or fundamental relation when combined with the SFR for field star-forming galaxies, these relationships suggest that part of the scatter in the M-Z relation is due to SFR. In order to test this scenario, we first look for any trend between the residuals of the M-Z relation and the SFR. As shown in Fig. \ref{fig:corrmetalsgf}, almost no correlation is observed between the two quantities ($\rho = -0.11$;  P($\rho$) $<$ 95\%). Attempting to correct the scatter in the M-Z relation using the SFR, based on a linear best fit to the residual oxygen abundance, actually increases the scatter in the M-Z relation from 0.11 dex to 0.13 dex. Using the NUV to estimate the SFR is reliant on the assumption that the SFR is constant over time-scales $>10^{8}$ yr, which may not be correct for cluster objects \citep{iglesias2004}. Therefore, we also performed the same test using SFRs calculated using \ha \ and FUV data (taken from \citealp{boselli2009}) available for 122 of the HRS galaxies. We briefly note that SFR(NUV) is typically greater than SFR(\ha) with a mean offset of 0.16 dex and a scatter of $\sigma$ = 0.4 dex. Once again, we found no trend between SFR and residual metallicity, nor could the scatter in the M-Z relation be reduced using a linear best fit to the SFR. This suggests the scatter in the M-Z relation is more closely related to the gas content than SFR. 

This is an interesting result, as the correlations between stellar mass, metallicity and gas fraction found in this work hint that the available gas content is more important than the SFR. A relation between metallicity and gas fraction is expected if a galaxy evolves like a closed box, without the inflow or outflow of gas (see e.g. \citealp{edmunds1990}). As an additional test, we investigate the $M_{*}-Z-\mu$ plane. In order to compare it's scatter to the scatter of the $M_{*}-Z-SFR$ plane, we follow the method of \citet{yates2011}. We do not find a minimum in the projection of least scatter for the SFR-plane (as in Fig. 6 of \citeauthor{yates2011}) nor do we find a significant minimum for the $\mu$-plane. This is most likely due to our sample being considerably smaller than that of \citeauthor{yates2011}, and we lack sufficient statistics to seek second order trends. When we attempt to express the stellar mass as a combination of the metallicity and SFR, using the method of \citet{laralopez2010}, we find an optimal fit to a plane which is consistent with their results and with a scatter of 0.12 dex. We note that the fact that we find a stronger correlation of metallicity with gas fraction than with SFR might have important implications. Future work will be to examine the role of the \hi \ content using a larger sample.

To summarise our main results so far, we find clear trends between the stellar mass, oxygen abundance and gas fraction, such that the scatter in the M-Z relation is anti-correlated with the gas content.

\section{On the effect of the environment}

The correlation between gas content and metallicity, whereby gas-poor galaxies are typically metal-rich, is consistent with an earlier study by \citet{skill1996}. They demonstrate that some individual galaxies within the cluster environment display metallicity enhancements of approximately 0.2 dex, whereas those galaxies on the periphery of the cluster have similar abundances as counterpart systems in sparse environments. In fact, many initial works based on small samples of galaxies (e.g. \citealp{ssk1991}; \citealp{hplc1992}; \citealp{hpc1994}; \citealp{skill1996}) suggest that the Virgo cluster members which typically display metallicity enhancements also have a tendency to be environmentally perturbed systems, as possibly indicated by a deficiency in their gas content. Recent studies using much larger samples have also indicated that gas deficient objects in clusters are more metal rich than field galaxies (\citealp*{boselli2006}). All these studies suggest that the gas content may affect the chemical abundance. Since \hi \ deficiency is typically linked to the removal of gas from a galaxy via environmental effects (e.g. ram pressure stripping), this could further indicate that the environment inhabited by a galaxy plays a role in chemical evolution processes. We thus proceed by investigating whether the environment inhabited by a galaxy effects the relations between stellar mass, metallicity and gas content, using the HRS cluster and field sub-samples. Since any difference in the metallicities of objects in different environments is likely to be small, as suggested by the 0.02 - 0.08 dex metallicity variations reported by \citet{mouhcine2007}, \citet{cooper2008}, and \citet{ellison2009}, we explore a number of different approaches to check for any variations with environment. 

\begin{figure*}[!tp]
\begin{center}
\includegraphics[width=0.75\textwidth]{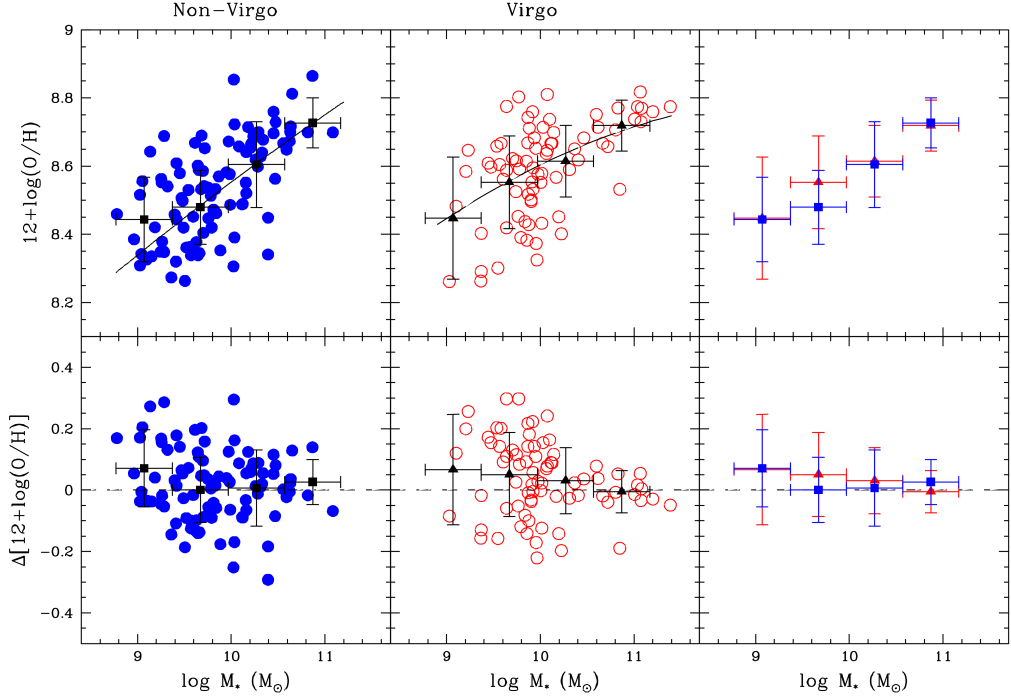}
\end{center}\vspace{-0.5cm}
\caption[The relationships between metallicity and gas fraction]{Stellar mass versus oxygen abundance (\textit{upper panels}) and the residual oxygen abundance (\textit{lower panels}) as calculated from the best-fit M-Z relation for the field sub-sample (solid black line). The sample is divided between field, non-Virgo galaxies (\textit{left panels}, solid blue circles) and those galaxies residing in the Virgo cluster (\textit{middle panels}, open red circles). Mean metallicities in bins of fixed stellar mass (\textit{right panels}) show no systematic difference between the field galaxies (blue squares) and Virgo cluster members (red triangles). Error bars show the 1-$\sigma$ scatter in each bin.}\label{fig:environmentmzr}
\end{figure*}

Firstly, we attempt to seek any difference between the scatter of the observations in the two different environments. Fig. \ref{fig:environmentmzr} presents M-Z relations for the Virgo cluster members and those galaxies residing outside the cluster.  The M-Z relation displays only minor differences between galaxies in the Virgo cluster and galaxies residing in sparser environments. From Fig. \ref{fig:environmentmzr}, the scatter yields no significant difference between the two distributions, as the dispersion is 0.15 dex in the cluster compared to a dispersion of 0.14 dex for the field galaxies. The correlation in the M-Z diagram of the cluster galaxies is also similar, with Spearman correlation coefficients of 0.48 in the cluster and 0.56 in the field. These coefficients relate to probabilities of the two quantities being correlated of  P($\rho$) $>$99\% . Thus, cluster galaxies do not display significant metallicity enhancements with respect to systems residing in the field. In addition, comparing the mean value in binned data for the cluster and field environments (right panel of Fig. \ref{fig:environmentmzr}), it is apparent that whilst the differences in the average metallicity estimates at each fixed stellar mass are of the order of 0.05 dex, there is no systematic metallicity enhancement across the range of mass bins observed for either environment. Indeed, neither environment shows an enhanced metallicity in any two adjacent mass bins. Finally, using the field M-Z relation to calculate the residual metallicities for galaxies in both environments (lower panels of Fig. \ref{fig:environmentmzr}), we further demonstrate there is no significant difference between the cluster and field M-Z relations. 

Overall, the Virgo cluster members do not demonstrate a higher dispersion compared to the non-cluster galaxies. No systematic variations between the environments are observed and no evidence suggests an environmental dependence to the scatter of the M-Z relation. However, we note that although it appears that the M-Z relation is insensitive to the environment a galaxy inhabits, we can not rule out the possibility that a subtle underlying variation is hidden within the observational errors. This may not be such a surprising result. Although \cite{ellison2009} found that the metallicity enhancement in the cluster M-Z relation was up to 0.05 dex, they also warn that such environmental differences are subtle and may not be clearly observed in the unbinned data of even large samples ($>$1300) of galaxies. Whilst the results are not sensitive enough to discriminate whether the small variations we observe are really due to the environment or merely arise from observational errors, we can place an upper limit of 0.1 dex on any environmental variation. This result is consistent with \cite{ellison2009} and \cite{mouhcine2007}, appearing to disagree with the observations of individual objects in \citet{skill1996}, who suggest that the environment plays a greater role in affecting the chemical abundance in comparison to our findings. Whilst it is possible that individual objects do have enhanced metallicities in the cluster environment, such strong environmental dependence is not statistically seen in any of the M-Z diagrams presented in Fig. \ref{fig:environmentmzr}.      

\begin{figure*}[!bp]
\begin{center}
\includegraphics[width=0.75\textwidth]{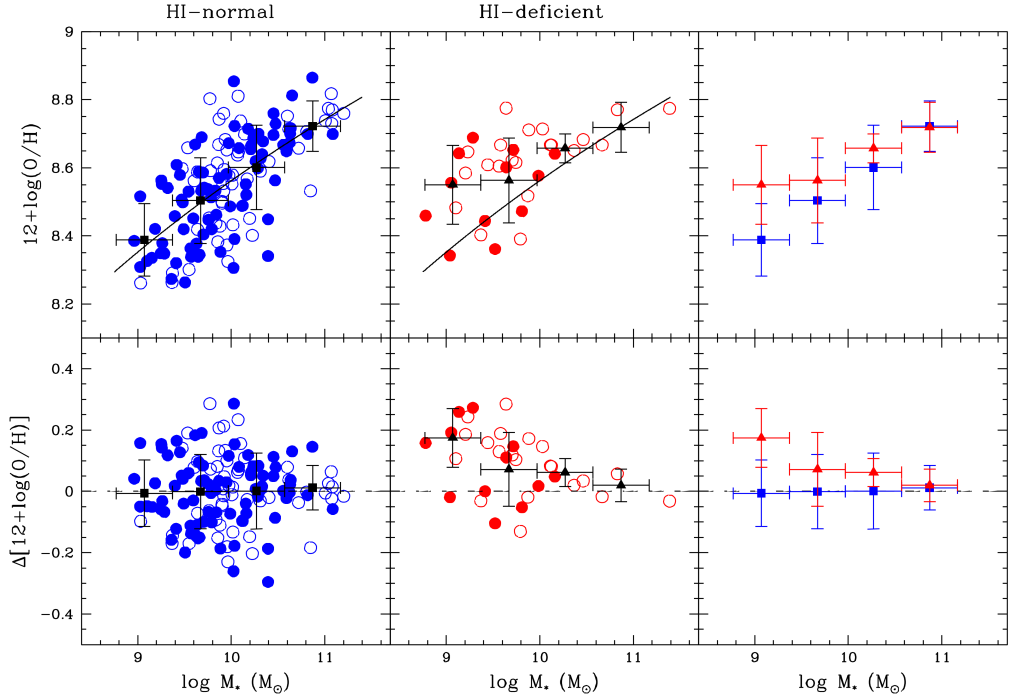}
\end{center}\vspace{-0.5cm}
\caption[The M-Z relations observed from galaxies residing in the Virgo cluster region and outside the cluster environment]{Stellar mass versus oxygen abundance (\textit{upper panels}) and the residual oxygen abundance (\textit{lower panels}) as calculated from the best-fit M-Z relation from the field sub-sample (solid black line). The sample is divided between \hi \ normal galaxies (\textit{left panels}, blue circles) and their \hi \ deficient counterparts (\textit{middle panels}, red circles). We differentiate between field galaxies and Virgo cluster members using solid and open symbols, respectively. Mean metallicities in bins of  fixed stellar mass (\textit{right panels}) show a clear systematic difference between the \hi \ normal (blue squares) and \hi \ deficient galaxies (red triangles). Error bars show the 1-$\sigma$ scatter in each bin.}\label{fig:hidefnew}
\end{figure*}

Initial inspection of the distribution of \hi -deficient galaxies on the M-Z diagram (upper left panel of Fig. \ref{fig:gasfrachidef}), appears to show that a large fraction (nearly $\sim$2/3) of the \hi \ deficient galaxies are metal-rich, i.e. they lie above the M-Z relation. In fact, by dividing the sample according to gas content (Fig. \ref{fig:hidefnew}), it is clear that \hi \ deficient galaxies typically display enhanced oxygen abundances compared to their \hi \ normal counterparts. The trend is clearer when using the average metallicities for each bin of fixed stellar mass, also shown in Fig. \ref{fig:hidefnew}, as this demonstrates that the \hi \ deficient galaxies do not follow the same M-Z relation derived from the \hi \ normal sample, but typically have enhanced metallicities. We also find marginal evidence that \hi \ deficiency seems to affect the metallicities of low mass galaxies more than high mass ones, as recently shown by \citet{petropoulou2012}. This may be due to high mass galaxies possessing flat metallicity gradients (\citealp{moran2012}), such that the removal of gas from the outer disk does not effect on the mean metallicity.

As \hi \ deficient objects are usually found in the cluster
environment, where infalling galaxies may be victim to the
effects of ram pressure stripping by the intracluster medium,
we might expect to observe a much stronger offset between
the cluster and field M-Z relations. However, considering that 1) only
$\sim$30\% (21 out of 75 objects) of the Virgo cluster members
in our sample are classed as \hi \ deficient, and 2) only 12 \hi \ deficient
Virgo galaxies display metal enhancements larger than 0.1 dex, it is not
entirely surprising that we do not observe average offsets larger than
0.05 dex between the field and cluster samples.
We only find a significant offset in the M-Z relation when we
divide our sample by \hi \ deficiency because we are preferentially selecting
galaxies that have already been significantly affected by their
environment. Nevertheless, we stress that larger samples of \clearpage

\clearpage

\noindent
cluster galaxies with integrated spectra are required to confirm our findings.

Our result that \hi \ deficient galaxies typically have enhanced metallicities is consistent with the properties of individual objects included in \citet{skill1996}, where some \hi \ deficient objects are more metal-rich with respect to those with normal gas content. Results from the Herschel Space Telescope have uncovered a correlation between the ratio of the submillimetre-to-optical diameter and with the \hi -deficiency, suggesting that the cluster environment is capable of stripping not only gas but also dust and, likely, metals \citep{cortese2010}. One possible explanation for a link between \hi \ deficiency and metallicity is that the removal of the outer gas-rich disk, via e.g. ram pressure stripping, would cut-off the inflow of metal-poor gas to the central regions. If a galaxies metallicity is the result of an equillibrium between the metal-poor inflow and SFR \citep{finlator2008}, then cutting gas inflow would cause an increase in metallicity as the galaxy continues to form stars. There also remains a small possibility that some metals avoid being stripped along with the gas and a fraction of the dust. The amount of dust removed from a galaxy has been demonstrated to be significantly lower than the amount of \hi \ stripped (\citealp{cortese2012dust}). Effective metal retention combined with efficient gas stripping could also produce metallicity enhancements for cluster galaxies \citep{petropoulou2012}.

\begin{figure*}[!t]
\begin{center}
\includegraphics[width=0.4\textwidth]{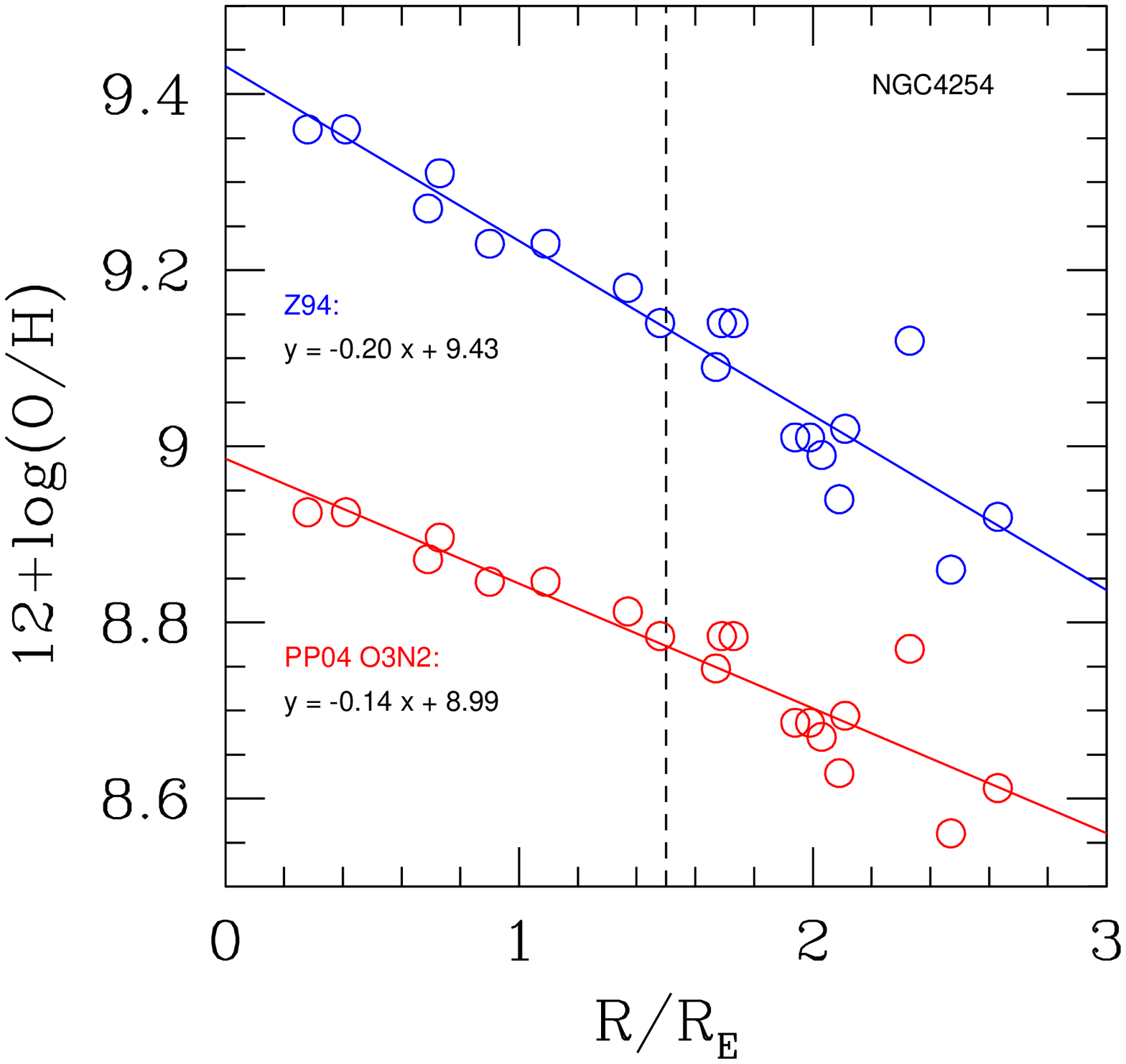}
\includegraphics[width=0.4\textwidth]{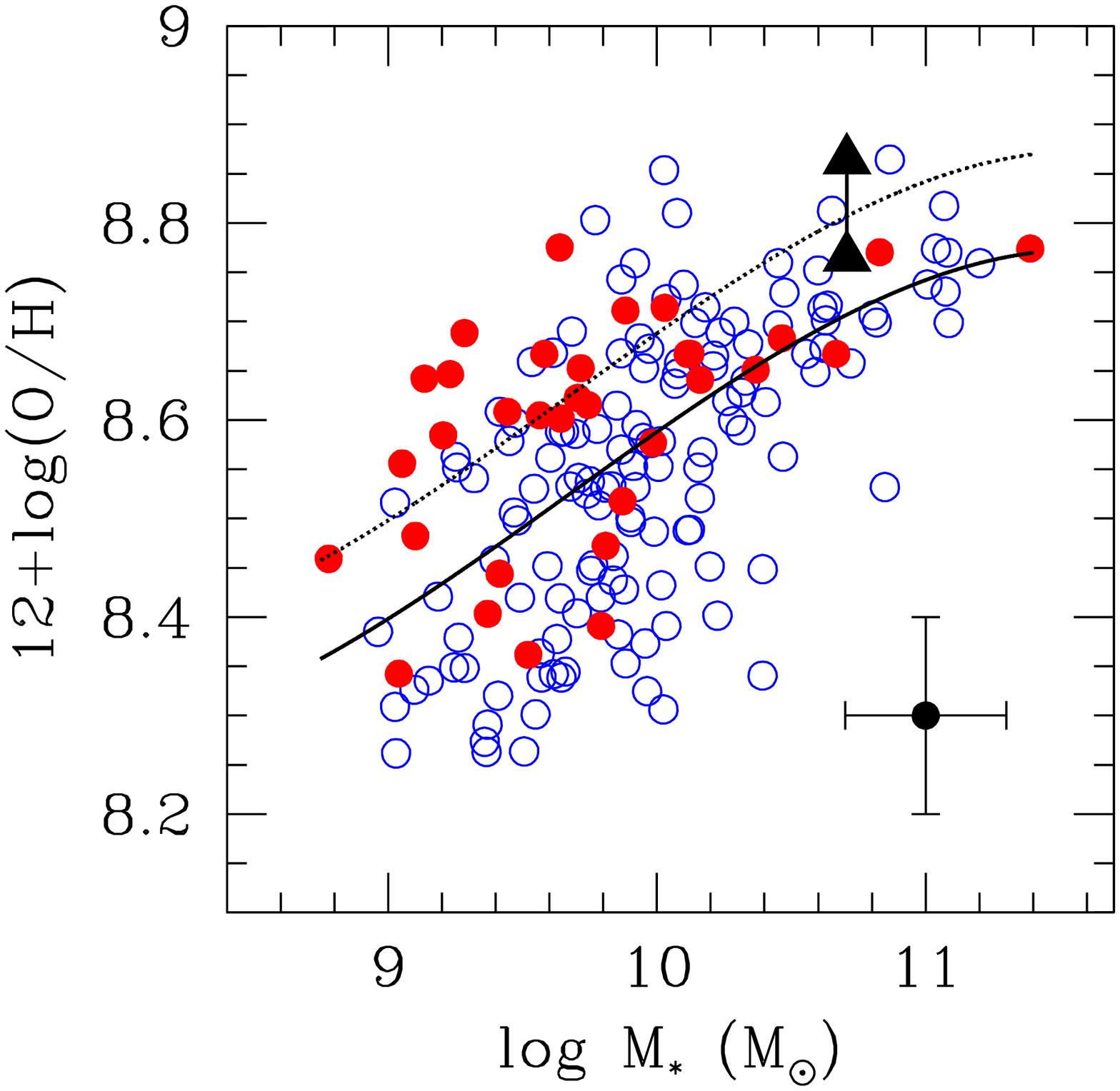}
\end{center}\vspace{-0.75cm}
\caption[A possible explanation for enhanced metallicities in \hi \ deficient galaxies]{\textit{Left panels}: The metallicity profile for NGC 4254 from \citet{skill1996} using the published Z94-based metallicities (blue circles) and converted into the PP04 O3N2 base calibration (red circles). The black dashed line marks the assumed stripping radius at $R/R_{E} < 1.5$. \textit{Right panels}: The two limits for NGC 4254 (black triangle) plotted on the M-Z diagram are based on the integrated metallicities for the total profile and for $R/R_{E} < 1.5$ i.e. `before' and `after' gas-stripping. The galaxy is shifted from the total M-Z relation (solid black line) to the best fit relation (dashed black line) to the \hi \ deficient galaxies (red circles).} \label{fig:hidefe}
\end{figure*}

An alternative explanation for the observed metallicity enhancements is based on the possibility of a selection effect in the observations. Consider the scenario where gas is stripped from the galaxy via an environmental mechanism, such as ram pressure stripping (see e.g. \citealp{boselli2006a}). In the outskirts of a galaxy, this will lead to a reduction in the number of observed \hii regions, since a reduction of the star formation rate following the removal of gas, combined with the effects of stellar evolution, means that the number of hot young stars emitting ionising photons will decrease. Therefore, only \hii regions within the stripping radius will be observed, where gas remains as fuel for new stars to continue to be created, and, in the case of galaxies with strong metallicity gradients, only the most metal rich \hii regions will contribute to an integrated measurement of the metallicity. In this scenario, a \hi \ deficient galaxy might appear to have an enhanced metallicity. To test this scenario, we use the radial abundance profile of NGC 4254 presented in \cite{skill1996}. This galaxy was arbitrarily chosen due to it's clear oxygen abundance gradient (when using the Z94 metallicity calibration). The Z94 profile is first converted into the PP04 O3N2 base calibration, using the same conversion relation as provided in KE08, such that integrated metallicities from the profile are comparable with the PP04 O3N2-based M-Z diagrams. The Z94 and converted PP04 profiles are presented in the left panel of Fig. \ref{fig:hidefe}.

Firstly, we note that the metallicity gradient is dependent on the choice of calibration used for the base metallicity, as shown by \citet{moustakas2010}. The Z94 profiles from \citet{skill1996} were chosen for displaying the strongest metallicity gradients of the nine galaxies in the \citet{skill1996} sample, with the aim of demonstrating the largest possible effect on the observed metallicity due to ram pressure stripping. Yet the metallicity gradient is very weak when, for example, the D02 acts as the base calibration. It is important to remember that the choice of calibration affects the observed metallicity gradient.

We obtained integrated metallicities observed in the cases `before' and `after' gas stripping. The integrated metallicity `before' gas stripping was found by weighting the metallicity gradient by an exponentially declining intensity profile $\propto e^{-R/R_{E}}$, and finding the average intensity-weighted metallicity for the full profile. To calculate the observed metallicity `after' gas stripping, the average intensity-weighted metallicity  was found within an assumed stripping radius of $R/R_{E} = 1.5$. This value is purely an approximation given that the stripping radius found in galaxy models varies depending on the model ingredients and are often around 11-15 kpc in a galaxy with semimajor axis of 25 kpc \citep{kron2008,tonne2009}. However, this assumption is sufficient for the illustrative purposes of this toy model. From these calculations, it is shown that metallicity enhancements of up to $\sim$0.1 dex may be reproduced, and which may be larger for galaxies with steeper metallicity gradients. This is a significant increase to the observed metallicity. The predicted metallicity enhancement moves NGC 4254 from the M-Z relation into the parameter space typical of the \hi \ deficient galaxies (Fig. \ref{fig:hidefe}). Thus, this simple model may explain the higher abundances of gas-deficient galaxies without requiring any special physical process that would alter the M-Z relation. Only 2D metallicity maps will be able to determine whether this increase
in metallicity is `real' or just due to an `observational bias'.  

To summarise the main points of this section, we find no significant variation of the M-Z relations interior or exterior to the Virgo cluster, but also cannot conclusively rule out the possibility of a weak sensitivity to the environment. We also demonstrated how a possible selection effect could enhance the observed abundances of \hi \ deficient galaxies without invoking an environmental variation in the M-Z relation. All these findings, consistent with previous studies, suggest that environmental processes play a secondary role in governing the M-Z relation, with internal processes playing a role in producing the observed relationships between stellar mass, gas content and metallicity.

\section{Discussion}\label{sec:discussion} 

In our analysis, we find that the gas content appears to drive the M-Z relation, given that we observe an anti-correlation between gas content and metal content, which remains even when removing the metallicity dependency on stellar mass. We also find no significant difference between M-Z relations of galaxies in cluster and field environments. Whilst the origin of the stellar mass - metallicity relation has been the focus of many previous studies, these new results are only made possible due to the unique dataset available for the HRS sample. 

Firstly, most previous studies have lacked spatially integrated, drift-scan spectra for a moderately-sized galaxy sample, instead using SDSS fiber spectroscopy or standard long-slit spectroscopy (see e.g. \citealp{garnett2002}; \citealp{tremonti2004}; \citealp{dalcanton2007}). Such data can prove problematic for the interpretation of results, since the presence of metallicity gradients may not be properly taken into account. Further complications may arise when using inhomogeneous metallicity estimates, either taken from the literature or calculated from multiple metallicity calibrations. Finally, and most importantly for this work, information on the gas content of galaxies is crucial for discriminating between the various scenarios that produce the M-Z relation; it is impossible to ascertain whether a galaxy is metal-poor because of the outflow of enriched gas, or a lower efficiency at converting gas into stars. Nearly all the previously mentioned studies lack direct gas measurements, instead inferring the gas content indirectly from the SFR with the Kennicutt relation (e.g. \citealp{tremonti2004}; \citealp{erb2008}; \citealp{spitoni2010}). Furthermore, previous studies examining the role of environment in shaping the M-Z relation also either lacked information on the gas content (\citealp{mouhcine2007}; \citealp{cooper2008}; \citealp{ellison2009}), or used small samples in their analysis (e.g. \citealp{skill1996}; \citealp{vila92}). Thus, this paper presents the first moderately large sample with spatially integrated, homogeneous metallicities derived from consistent metallicity calibrations, \hi \ gas measurements and environment information. Our findings that the gas content drives the M-Z relation could not be done without this information. 

\begin{figure*}[!t]
\begin{center}
\includegraphics[width=0.32\textwidth]{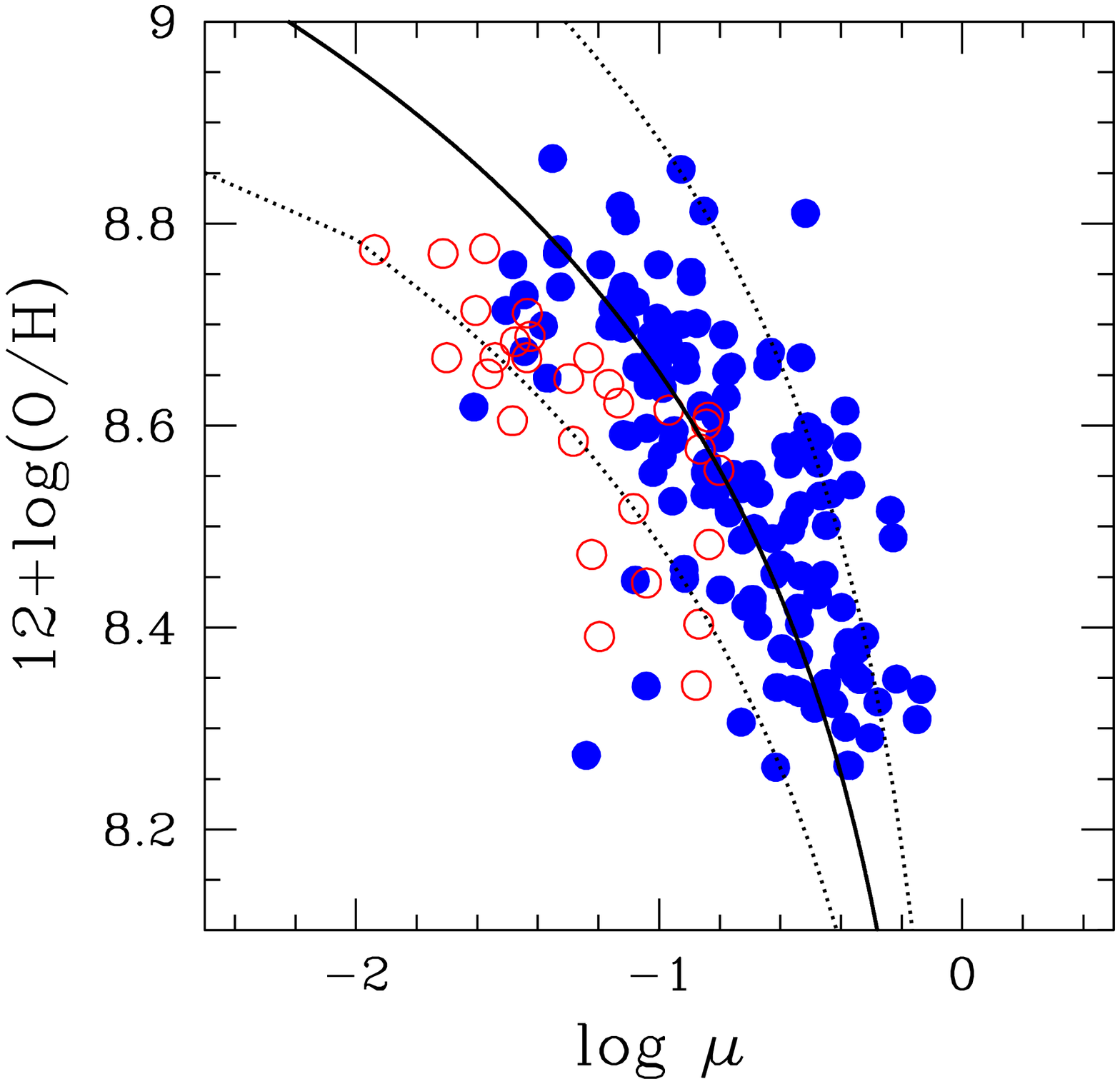}
\includegraphics[width=0.32\textwidth]{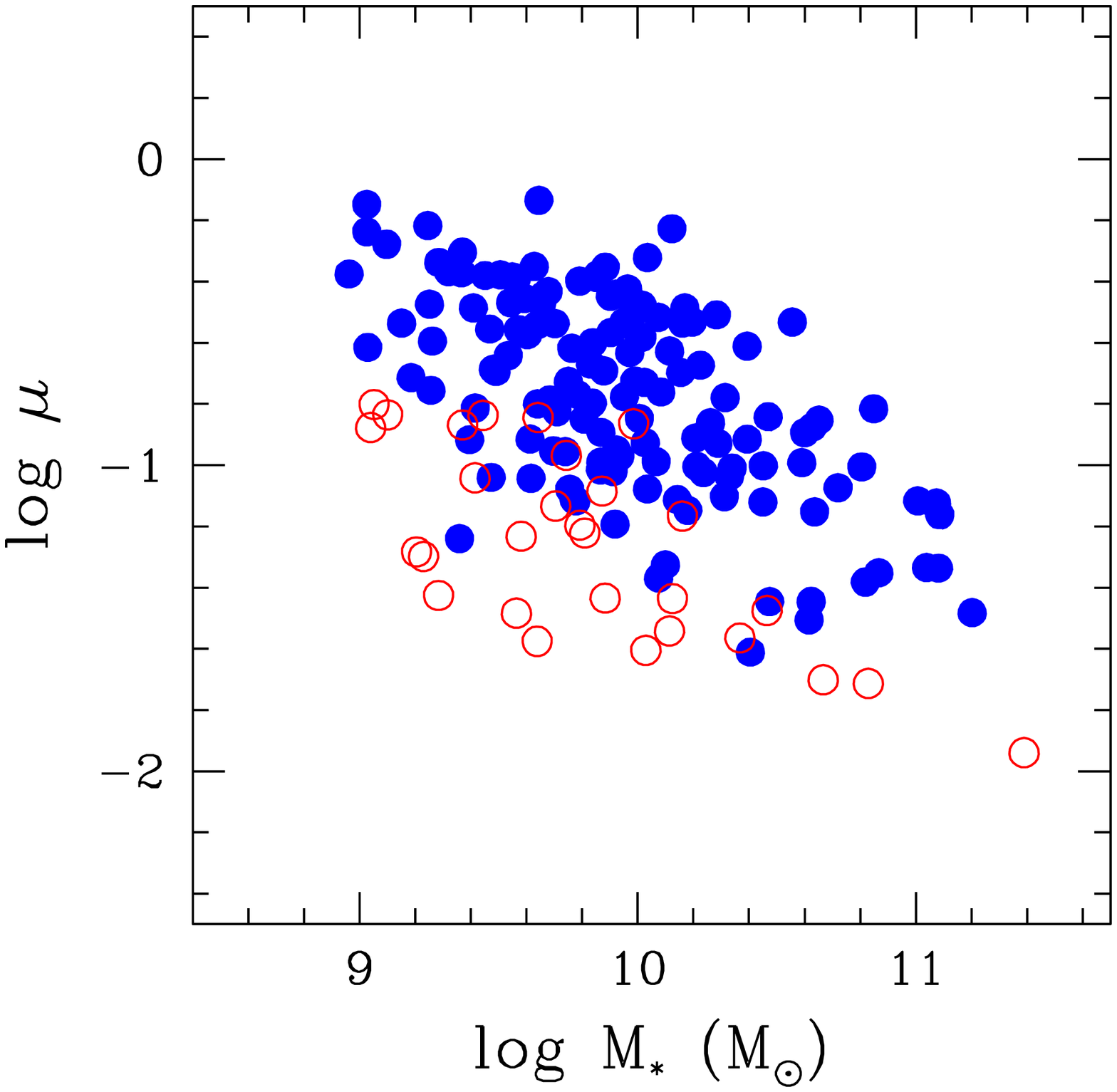}
\includegraphics[width=0.32\textwidth]{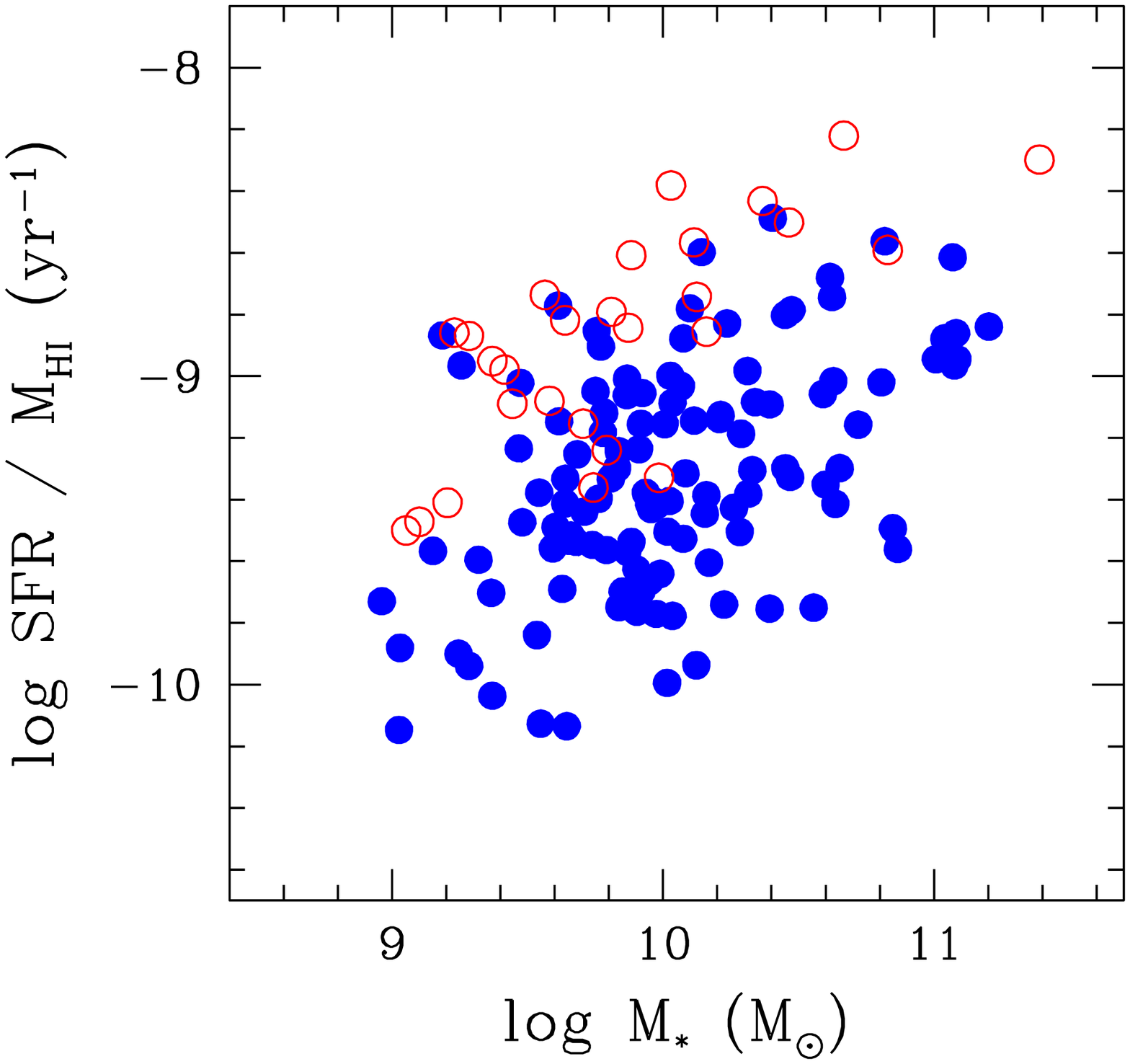}
\end{center}\vspace{-0.75cm}
\caption[The oxygen abundance versus the logarithm of the gas fraction, fitted with the prediction of the closed-box model]{\textit{Left}:The oxygen abundance versus the logarithm of the gas fraction from observed galaxies compared with the prediction of the closed-box model (solid black line) together with the 1$\sigma$ limits (dashed black line). \textit{Middle}: The logarithm of the gas fraction versus the stellar mass. \textit{Right}: The ratio of the SFR(NUV) to the mass of \hi \ versus the stellar mass. In all three plots, \hi \ normal and \hi \ deficient galaxies are plotted as solid blue and open red circles, respectively.}\label{fig:gfpeff}
\end{figure*}

As we noted earlier, the relationship between gas content and metallicity is expected if a galaxy evolves like a closed box, without the inflow or outflow of gas (see e.g. \citealp{edmunds1990}). Even though galaxies are thought to form heirarchically (e.g. \citealp{springel2005}; \citealp{baugh2006}) and evidence shows that galaxies do not evolve as truely closed systems (e.g. \citealp*{boselli2006}), it is still important to test that these results are indeed consistent with the simple closed box model. The model predicts that as star formation converts gas into stars, the gas metallicity $Z_{gas}$ increases as the gas mass fraction $\mu$ decreases according to \citep{ss72}
\begin{eqnarray}\label{eq:peff}
Z_{gas} \equiv  y \ln (\mu^{-1}) ,
\end{eqnarray}
where $y$ is the true nucleosynthetic yield, defined as the mass in heavy elements freshly produced by massive stars and returned to the ISM relative to the total mass locked up in long-lived stars and stellar remnants. In reality, a galaxy can exchange mass with it's environment, which will alter the above relation and mimic a variation in the oxygen yield. The simple closed box model can then be used to estimate the effective yield of oxygen, $y_{eff}$. We briefly note that $y_{eff}$ should be constant for closed box evolution. In Fig. \ref{fig:gfpeff} (left panel), we compare the observed relationship between gas content and metallicity with the prediction from the closed-box model of chemical evolution, adopting the observed average effective yield of $y_{eff} =  10^{-2.6}$, consistent with \citet{pilyugin2004} and \citet{pilyugin2007}. Though the mean trend of the galaxies follows the prediction of the closed box model, there is obvious scatter in the \hi \ normal galaxies. This is likely due to our inability to take into account the molecular hydrogen in the gas fraction and also with our choice of IMF used to calculate stellar masses; switching the Salpter (1955) IMF with a Kroupa (2001) IMF would lower our stellar masses by 0.2 dex. Additionally, some galaxies are not well represented by the observed average effective yield. In a study of the local M-Z relation for 53,000 local galaxies, \citet{tremonti2004} find an observed correlation between baryonic mass and effective yield (see their Fig. 8), and interpret the shape of the relation in terms of efficient galactic winds that remove metals from low-mass galaxies ($\leq$ 10$^{10.5}$ M$_{\odot}$). We note that we do not observe the same trend between baryonic mass and effective yield, most likely because our sample does not span a large enough range in stellar mass, and we instead find a constant average effective yield of $y_{eff} =  10^{-2.6}$ with a 0.2 dex scatter across the mass range. Our results are consistent with the predictions
of a closed-box model. However, \citet{dalcanton2007} showed analytically that any change to the effective yield due to gas flows may quickly be returned to the true yield expected by closed-box evolution. Therefore, it may not be possible to draw conclusions about closed-box evolution or the impact of inflows or outflows by using  measurements of the effective yields alone.

Despite this, our results apparently support a scenario in which the M-Z relation is a consequence of a variation in the star formation efficiency as a function of stellar mass (see e.g., \citealp{brooks2007}). The idea that less massive systems are less evolved compared to larger galaxies, due to the efficiency of star formation being larger in more massive systems, has both observational (e.g \citealp{leq1979}; \citealp{matteucci1994}; \citealp{boselli2001}) and theoretical support from N-body simulations (e.g. \citealp{mouhcine2008}). Our results are most consistent with this scenario; more massive galaxies typically have lower gas content (see the middle panel of Fig. \ref{fig:gfpeff}) and higher metallicities (Fig. \ref{fig:mzrshighlighted}), whereas less massive galaxies are still gas-rich and metal-poor, suggesting that the lower mass systems are less efficient at converting gas into stars which produce metals. In addition, the ratio of the star formation rate to the gas mass, which provides a measure of the galaxies current SF efficiency, correlates with M$_{*}$ with a Spearman correlation coefficient of $\rho$ = 0.41 (P($\rho$) $>$ 99.9 \%). The fact that more massive galaxies have lower gas fractions and higher present-day SF efficiencies (right panel of Fig. \ref{fig:gfpeff}) compared to lower mass systems is consistent with the scenario where the M-Z relation arises from a mass-dependent variation in the star formation efficiency. Thus, as a galaxy evolves, these mass-dependent variations in SF efficiency govern the conversion of gas into stars, which produce the observed relations between gas content, stellar mass and metallicity.

For completeness in this discussion, we also note that a variable integrated stellar initial mass function may also give rise to the M-Z relation (\citealp{koppen2007}). Whether the IMF is universal for all environments has been the focus of recent debate; a few studies have found indirect observational evidence that the IMF may not be universal and instead vary as a function of stellar mass, star formation activity or gas column density (e.g \citealp{lee2004}; \citealp{hoversten2008}; \citealp{meurer2009}). However, more direct approaches to measure the IMF in the Milky Way and local galaxies support a universal IMF (e.g. \citealp{massey1995}; \citealp{scalo1998}; \citealp{selman2008}). In a recent study, we investigated the possibility of a variable IMF by studying the high mass star formation activity of the HRS late-type galaxies, using the \ha \ and the FUV luminosity as two independent, direct tracers of star formation (\citealp{boselli2009}). Our results are consistent with a \citet{kroupa2001} and \citet{salpeter1955} IMF in the high mass stellar range ($>$ 2 M$_{\odot}$), and we show that slight variations in the IMF slope with mass can be due to the different micro-histories of star formation in massive galaxies with respect to dwarf systems and does not require a variable IMF. However, at present we cannot exclude the possibility of a variable IMF, since we still do not have a clear picture of whether or not the function is universal (see e.g. \citealp{meurer2011}). 

In an analytical study of the possible physical mechanisms that could contribute towards the M-Z relation, \cite{spitoni2010} tested the cases of inflows and outflows of gas, variable gas flow rates, and the variable IMF scenario proposed by \citet{koppen2007}. They conclude that whilst galactic winds and the variable IMF cannot be excluded as possible explanations for the M-Z relation, the best solution could be a variable efficiency of the star formation rate with a possible effect of out-flowing gas in lower mass galaxies. Therefore, whilst at present it is difficult to identify one particular mechanism giving rise to the M-Z relation using the HRS sample, we find the most likely interpretation is for a varying star formation efficiency between high and low mass systems. In this scenario, the M-Z relation arises out low mass galaxies being less efficient at converting gas into metals compared to higher mass systems. Unfortunately, it is not possible to analytically explore the case of increasing star formation efficiency with increasing mass, since the SFR does not appear in the solution of the closed-box model derived above. This scenario has however recently been studied by \citet{calura2009}, who find that the M-Z relation can be reproduced using an increasing efficiency of star formation with mass in galaxies of all morphological types, without any need to invoke inflows of pristine gas or outflows of enriched gas that favour the loss of metals in the less massive galaxies. Their findings successfully predict the M-Z relation not just in the local universe, as studied in this work, but also out to the high redshift universe.  

We note that explaining the correlation between stellar mass and gas content still poses a challenge for theoretical models, as the correlation must be tight in order to ensure a low scatter in the M-Z relation. Using cosmological hydrodynamic simulations, \citet{dave2011} explore the effects of inflows, outflows and star formation in governing the gas and metal content. At $z=0$, all their wind models predict a declining gas fraction with increasing stellar mass for higher mass systems, in agreement with observations. However, at lower masses this trend shows a turnover to lower gas fractions with decreasing stellar mass. Such behavior is not seen in our observations (see also \citealp{cortese2011}), nor in those from past studies (e.g. \citealp*{peeples2011}). Thus, accurately explaining the M-$\mu$ relation remains an important goal for future work, as this will provide a constraint on the underlying physical processes governing the stellar, gas and metal content. Since the dataset presented in this paper will be useful for constraining models in future studies on chemical evolution, we therefore provide for the community the derived physical quantities of the 169 HRS late-types that have metallicity estimates (see Table 3).

\section{Conclusions}

The aim of this work was to study the relationships between the stellar mass, metallicity, and gas content of galaxies in different environments. We used new optical, drift-scan spectroscopic observations of the HRS galaxy sample to obtain reliable metallicity estimates by taking the average oxygen abundances from five calibration methods. These measurements were combined with ultraviolet to near-infrared photometry and \hi \ 21 cm line observations to provide a multi-wavelength dataset capable of tracing the stellar, gas and metal content. 

We first demonstrated the reliability of our metallicity estimates and found further consistency between the stellar mass - metallicity relation observed in the HRS sample and those relations found in previous studies. A correlation was observed between the metal content and gas content of a galaxy, whereby gas poor galaxies are typically metal rich. We have shown that the removal of gas from the outskirts of spirals increases the observed metallicity by  $\sim$0.1 dex. We investigated whether any environmental variation in the shape or the scatter in the relationships could be present in the HRS sample, by looking at the properties of galaxies interior and exterior to the Virgo cluster. Although some cluster galaxies are gas-deficient objects, statistically the stellar-mass metallicity relation is nearly invariant to the environment. Although we cannot rule out the weak environmental trends reported by some recent studies (e.g. \citealp{mouhcine2007}; \citealp{ellison2009}), we conclude that any contribution by the environment is probably a secondary effect and that the relations are most likely driven by internal processes. Whilst at present it is difficult to identify one particular mechanism giving rise to the M-Z relation, we find the most likely interpretation of our results is that the M-Z relation originates from a varying star formation efficiency between high and low mass systems. Higher mass systems are able to convert their gas into stars more efficiently, producing a lower gas content and higher metal content with respect to lower mass galaxies.

The results presented here utilise PP04 O3N2 calibration as the base metallicity; we note that the choice of calibrator to act as the base metallicity does not affect the results or conclusions. Often, only a systematic shift was observed between multiple sets of results derived using different calibrations as the base metallicity, which is expected due to the relationship between the different calibrations. Since the same conclusions are reached independently of the choice of calibration used, it is likely that these conclusions are real and not spurious trends introduced via the method of metallicity estimation. Finally, these results are also based on relative measurements of the mean global oxygen abundance. Fluctuations in the temperature and density structures throughout \hii regions mean that even the ratio of collisionally-excited emission lines may not be reliable. It is possible that studies into diagnostic lines that are insensitive to the temperature and density, such as metal recombination lines (see e.g. \citealp{tsamis2003}; \citealp{liu2002}), or IR fine structure lines (e.g. \citealp{hunt2010}), may help resolve the metallicity discrepancy problem in the near future.

\begin{acknowledgements}
We gratefully acknowledge the constructive comments from the anonymous referee, which significantly improved the quality of this paper. TMH acknowledges the support of a Kavli Fellowship. The research leading to these results has received funding from the European Community's Seventh Framework Programme (/FP7/2007-2013/) under grant agreement No. 229517. This publication makes use of data from 2MASS, which is a joint project of the University of Massachusetts and the IPAC/Caltech, funded by the NASA and the NSF, and from the GALEX mission, developed in cooperation with the CNES-France and the Korean Ministry of Science and Technology. This research has made use of the NASA/IPAC Extragalactic Database (NED), which is operated by the Jet Propulsion Laboratory, California Institute of Technology, under contract with the National Aeronautics and Space Administration. This research has made use of the GOLDMine Database \citep{gavazzi2003}.
\end{acknowledgements}

\bibliography{mzrelation}
\onecolumn
\begin{landscape}
\LTcapwidth=\textwidth\begin{center}\tiny 
\begin{longtable}{l c l l c c c c c c c c c c c c c c c c c c c}
\caption[Oxygen abundances based on different line ratio calibration methods]{Oxygen abundance measurements for late-type galaxies. Column 1: galaxy designation.; Column 2: HRS designation.; Column 3,4: J2000 celestial coordinates.; Column 5: Balmer decrement.; Column 6: [\oii] flag, `.'/`$<$' for detected/undetected.; Column 7-11: Oxygen abundances from five different calibrations in units of 12 + log (O/H).; Column 12-16: Oxygen abundances converted from five different calibrations converted into the PP04 O3N2 base metallicity calibration, following the relations in KE08, in units of 12+log(O/H); Column 17: Mean metallicity.; Column 18: Error in metallicity.; Column 19: Number of calibrations used in the mean metallicity estimation.; Column 20: Reference for spectroscopy: 1 - \citeauthor{boselliprep} (2012), 2 - \citet{gavazzi2004}, 3 - \citet{moustakas2010}, 4 - \citet{moustakas2006}, 5 - \citet{jansen2000}, 6 - \citet{kennicutt1992}. See Section \ref{sec:metestimates} for details.}\label{tab:lines}\\
\hline \hline \\[-2ex]
\multicolumn{1}{c}{ } &
\multicolumn{1}{c}{ } &
\multicolumn{1}{c}{ } &
\multicolumn{1}{c}{ } &
\multicolumn{1}{c}{ } &
\multicolumn{1}{c}{ } &
\multicolumn{1}{c}{ } &
\multicolumn{5}{c}{Z} &
\multicolumn{1}{c}{ } &
\multicolumn{5}{c}{Z(PP04 O3N2)} &
\multicolumn{1}{c}{ } &
\multicolumn{1}{c}{ } &
\multicolumn{1}{c}{ } &
\multicolumn{1}{c}{ } &
\multicolumn{1}{c}{ } \\ \cline{8-12} \cline{14-18} 
\multicolumn{1}{c}{Galaxy} &
\multicolumn{1}{c}{HRS} &
\multicolumn{1}{c}{R.A.(h m s)} &
\multicolumn{1}{c}{Dec.($^{\circ \, \prime \, \prime \prime}$)} &
\multicolumn{1}{c}{C(\hb)} &
\multicolumn{1}{c}{[\oii]} &
\multicolumn{1}{c}{\hspace{1 mm} } &
\multicolumn{1}{c}{M91} &
\multicolumn{1}{c}{Z94} &
\multicolumn{1}{c}{KD02} &
\multicolumn{1}{c}{O3N2} &
\multicolumn{1}{c}{N2} &
\multicolumn{1}{c}{\hspace{1 mm} } &
\multicolumn{1}{c}{M91} &
\multicolumn{1}{c}{Z94} &
\multicolumn{1}{c}{KD02} &
\multicolumn{1}{c}{O3N2} &
\multicolumn{1}{c}{N2} &
\multicolumn{1}{c}{\hspace{1 mm} } &
\multicolumn{1}{c}{$\bar{Z}$} &
\multicolumn{1}{c}{E($\bar{Z}$)} &
\multicolumn{1}{c}{Methods} &
\multicolumn{1}{c}{Ref.} \\
\multicolumn{1}{c}{(1)} &
\multicolumn{1}{c}{(2)} &
\multicolumn{1}{c}{(3)} &
\multicolumn{1}{c}{(4)} &
\multicolumn{1}{c}{(5)} &
\multicolumn{1}{c}{(6)} &
\multicolumn{1}{c}{\hspace{1 mm} } &
\multicolumn{1}{c}{(7)} &
\multicolumn{1}{c}{(8)} &
\multicolumn{1}{c}{(9)} &
\multicolumn{1}{c}{(10)} &
\multicolumn{1}{c}{(11)} &
\multicolumn{1}{c}{\hspace{1 mm} } &
\multicolumn{1}{c}{(12)} &
\multicolumn{1}{c}{(13)} &
\multicolumn{1}{c}{(14)} &
\multicolumn{1}{c}{(15)} &
\multicolumn{1}{c}{(16)} &
\multicolumn{1}{c}{\hspace{1 mm} } &
\multicolumn{1}{c}{(17)} &
\multicolumn{1}{c}{(18)} &
\multicolumn{1}{c}{(19)} &
\multicolumn{1}{c}{(20)} \\[0.5ex] \hline
   \\[-1.8ex]
\endfirsthead
\multicolumn{23}{c}{{\tablename} \thetable{} -- Continued} \\[0.5ex]
  \hline \hline \\[-2ex]
\multicolumn{1}{c}{ } &
\multicolumn{1}{c}{ } &
\multicolumn{1}{c}{ } &
\multicolumn{1}{c}{ } &
\multicolumn{1}{c}{ } &
\multicolumn{1}{c}{ } &
\multicolumn{1}{c}{ } &
\multicolumn{5}{c}{Z} &
\multicolumn{1}{c}{ } &
\multicolumn{5}{c}{Z(PP04 O3N2)} &
\multicolumn{1}{c}{ } &
\multicolumn{1}{c}{ } &
\multicolumn{1}{c}{ } &
\multicolumn{1}{c}{ } &
\multicolumn{1}{c}{ } \\ \cline{8-12} \cline{14-18} 
\multicolumn{1}{c}{Galaxy} &
\multicolumn{1}{c}{HRS} &
\multicolumn{1}{c}{R.A.(h m s)} &
\multicolumn{1}{c}{Dec.($^{\circ \, \prime \, \prime \prime}$)} &
\multicolumn{1}{c}{C(\hb)} &
\multicolumn{1}{c}{[\oii]} &
\multicolumn{1}{c}{\hspace{1 mm} } &
\multicolumn{1}{c}{M91} &
\multicolumn{1}{c}{Z94} &
\multicolumn{1}{c}{KD02} &
\multicolumn{1}{c}{O3N2} &
\multicolumn{1}{c}{N2} &
\multicolumn{1}{c}{\hspace{1 mm} } &
\multicolumn{1}{c}{M91} &
\multicolumn{1}{c}{Z94} &
\multicolumn{1}{c}{KD02} &
\multicolumn{1}{c}{O3N2} &
\multicolumn{1}{c}{N2} &
\multicolumn{1}{c}{\hspace{1 mm} } &
\multicolumn{1}{c}{$\bar{Z}$} &
\multicolumn{1}{c}{E($\bar{Z}$)} &
\multicolumn{1}{c}{Methods} &
\multicolumn{1}{c}{Ref.} \\
\multicolumn{1}{c}{(1)} &
\multicolumn{1}{c}{(2)} &
\multicolumn{1}{c}{(3)} &
\multicolumn{1}{c}{(4)} &
\multicolumn{1}{c}{(5)} &
\multicolumn{1}{c}{(6)} &
\multicolumn{1}{c}{\hspace{1 mm} } &
\multicolumn{1}{c}{(7)} &
\multicolumn{1}{c}{(8)} &
\multicolumn{1}{c}{(9)} &
\multicolumn{1}{c}{(10)} &
\multicolumn{1}{c}{(11)} &
\multicolumn{1}{c}{\hspace{1 mm} } &
\multicolumn{1}{c}{(12)} &
\multicolumn{1}{c}{(13)} &
\multicolumn{1}{c}{(14)} &
\multicolumn{1}{c}{(15)} &
\multicolumn{1}{c}{(16)} &
\multicolumn{1}{c}{\hspace{1 mm} } &
\multicolumn{1}{c}{(17)} &
\multicolumn{1}{c}{(18)} &
\multicolumn{1}{c}{(19)} &
\multicolumn{1}{c}{(20)} \\[0.5ex]  \hline
   \\[-1.8ex]
\endhead 
 \\[-1.8ex] \hline
  \multicolumn{23}{c}{{Continued on next page\ldots}} \\
\endfoot

  \\[-1.8ex] \hline \hline
\endlastfoot
  CGCG 522-038  &  -   &      01 52 45.7  &  +36 37 07      &  1.10       &  $<$ & \hspace{1 mm}  &    -      &      -        &       -      &    -	 &    8.71 & \hspace{1 mm}   &    -	  &    -     &    -	&    -      &  8.78 & \hspace{1 mm} 	& 8.78   &   0.10  &   1     &   2  \\
  CGCG 522-041  &  -   &      01 52 53.7  &  +36 03 12      &  0.83      &  $<$  & \hspace{1 mm} &    -      &      -	     &        -      &   8.69	 &    8.62 & \hspace{1 mm}   &    -	  &    -     &    -	&   8.69    &  8.68 & \hspace{1 mm} 	& 8.68   &   0.01  &   2     &   2  \\
  CGCG 522-058  &  -   &      01 54 53.9  &  +36 55 05      &  0.90       &  $<$ & \hspace{1 mm}   &    -      &      -	     &        -      &    -	 &     -   & \hspace{1 mm}   &    -	  &    -     &    -	&    -      &	-   & \hspace{1 mm} 	&  -	 &    -    &   -     &   2  \\
\end{longtable}
\end{center}

\end{landscape}
\LTcapwidth=\textwidth\begin{center}\tiny \begin{longtable}{  c c c c c c c c}
\caption[Oxygen abundances based on different line ratio calibration methods]{Physical properties of the HRS late-type galaxies. Column 1: HRS designation.; Column 2: Logarithm of stellar mass, in M$_{\odot}$ units.; Column 3: Logarithm of \hi \ mass in M$_{\odot}$ units.; Column 4: Reference for HI data (see footnote).; Column 5: \hi \ deficiency.; Column 6: Oxygen abundance derived in this work in units of 12+log(O/H).; Column 7: SFR based on GALEX NUV data.; Column 8: Galaxy environment: V - Virgo cluster member, F - field/non-Virgo system.}\label{tab:modlines}\\
\hline \hline \\[-2ex]
\multicolumn{1}{c}{HRS} &
\multicolumn{1}{c}{$M_{*}$} &
\multicolumn{1}{c}{$M_{HI}$} &
\multicolumn{1}{c}{Ref. (HI)} &
\multicolumn{1}{c}{DEF(HI)} &
\multicolumn{1}{c}{Z} &
\multicolumn{1}{c}{SFR(NUV)} &
\multicolumn{1}{c}{Env.} \\
\multicolumn{1}{c}{ } &
\multicolumn{1}{c}{M$_{\odot}$} &
\multicolumn{1}{c}{M$_{\odot}$} &
\multicolumn{1}{c}{ } &
\multicolumn{1}{c}{ } &
\multicolumn{1}{c}{12+log(O/H)} &
\multicolumn{1}{c}{M$_{\odot}$ yr$^{-1}$} &
\multicolumn{1}{c}{ }\\
\multicolumn{1}{c}{(1)} &
\multicolumn{1}{c}{(2)} &
\multicolumn{1}{c}{(3)} &
\multicolumn{1}{c}{(4)} &
\multicolumn{1}{c}{(5)} &
\multicolumn{1}{c}{(6)} &
\multicolumn{1}{c}{(7)} &
\multicolumn{1}{c}{(8)}\\[0.5ex] \hline
   \\[-1.8ex]
\endfirsthead
\multicolumn{8}{c}{{\tablename} \thetable{} -- Continued} \\[0.5ex]
  \hline \hline \\[-2ex]
\multicolumn{1}{c}{HRS} &
\multicolumn{1}{c}{$M_{*}$} &
\multicolumn{1}{c}{$M_{HI}$} &
\multicolumn{1}{c}{Ref. (HI)} &
\multicolumn{1}{c}{DEF(HI)} &
\multicolumn{1}{c}{Z} &
\multicolumn{1}{c}{SFR(NUV)} &
\multicolumn{1}{c}{Env.} \\
\multicolumn{1}{c}{ } &
\multicolumn{1}{c}{M$_{\odot}$} &
\multicolumn{1}{c}{M$_{\odot}$} &
\multicolumn{1}{c}{ } &
\multicolumn{1}{c}{ } &
\multicolumn{1}{c}{12+log(O/H)} &
\multicolumn{1}{c}{M$_{\odot}$ yr$^{-1}$} &
\multicolumn{1}{c}{ }\\
\multicolumn{1}{c}{(1)} &
\multicolumn{1}{c}{(2)} &
\multicolumn{1}{c}{(3)} &
\multicolumn{1}{c}{(4)} &
\multicolumn{1}{c}{(5)} &
\multicolumn{1}{c}{(6)} &
\multicolumn{1}{c}{(7)} &
\multicolumn{1}{c}{(8)} \\[0.5ex] \hline
   \\[-1.8ex]
\endhead 
 \\[-1.8ex] \hline
  \multicolumn{8}{c}{{Continued on next page\ldots}} \\
\endfoot

  \\[-1.8ex] \hline \hline
\endlastfoot
  1   &   9.13   &     -    &  -    &	  -        &  8.64$\pm$0.11  &   -    &   F   \\
  2   &   9.25   &    8.45  &  1    &	 0.04	   &  8.55$\pm$0.19  &  0.39  &   F   \\
  10  &   9.48   &    8.76  &  2    &	 0.12	   &  8.49$\pm$0.09  &   -    &   F   \\
  11  &   9.86   &    8.79  &  3    &	 0.24	   &  8.57$\pm$0.09  &  0.85  &   F   \\
  12  &   8.96   &    8.69  &  4    &	 -0.4	   &  8.38$\pm$0.10  &  0.14  &   F   \\
  13  &   10.81  &    9.32  &  3    &	 0.25	   &  8.69$\pm$0.05  &   -    &   F   \\
  16  &   10.23  &    9.12  &  5    &	 0.03	   &  8.68$\pm$0.01  &  1.08  &   F   \\
  17  &   10.15  &    9.42  &  3    &	 -0.09     &  8.55$\pm$0.06  &  1.62  &   F   \\
\end{longtable}
\end{center}
{\scriptsize 
References: 
(1) \cite{bicay96_2}
(2) \cite{peterson79}
(3) \cite{springob2005}
(4) \cite{rosenberg00}
(5) \cite{hewitt83}
(6) \cite{davis83}
(7) \cite{stierwalt09}
(8) \cite{giovanelli2007}
(9) \cite{kent2008}
(10) \cite{bicay96_1}
(11) \cite{bicay87}
(12) \cite{smith88}
(13) \cite{helou82}
(14) \cite{oneil04}
(15) \cite{schneider92}
(16) \cite{courtois09}
(17) \cite{richter87}
(18) \cite{huchtmeier85}
(19) \cite{huchtmeier05}
(20) \cite{helou84}
(21) \cite{lu93}
(22) \cite{bottinelli82}
(23) \cite{smoker00}
(24) \cite{hoffman95}
(25) \cite{hoffman89}
(26) \cite{fisher81}
(27) \cite{schneider90}
(28) \cite{hoffman89b}
(29) \cite{bottinelli90}
(30) \cite{hoffman87}
(31) \cite{koribalski04}
(32) \cite{haynes79}
(33) \cite{gavazzi06}
(34) \cite{theureau98}
(35) \cite{rc3}
(36) \cite{haynes81}
(37) \cite{lewis85}
(38) \cite{huchtmeier89}
}

\end{document}